\documentclass[english]{article}
\usepackage[T1]{fontenc}
\usepackage[latin9]{inputenc}
\usepackage{geometry}
\usepackage{color}
\usepackage{amsmath}
\usepackage{amssymb}
\usepackage{esint}

\makeatletter
\newcommand{\lyxaddress}[1]{
\par {\raggedright #1
\vspace{1.4em}
\noindent\par}
}

\makeatother

\usepackage{babel}
\begin{document}

\title{Understanding the Pointer States}

\author{Carlos Alexandre Brasil}

\maketitle

\lyxaddress{carlosbrasil.physics@gmail.com Instituto de Física \textquotedbl{}Gleb
Wataghin\textquotedbl{}, Universidade Estadual de Campinas, P.O. Box
6165, 13083-970 Campinas, SP, Brazil }

\author{Leonardo Andreta de Castro}

São Carlos Institute of Physics, University of São Paulo, PO Box 369,
13560-970, São Carlos, SP, Brazil
\begin{abstract}
In quantum mechanics, pointer states are eigenstates of the observable
of the measurement apparatus that represent the possible positions
of the display pointer of the equipment. The origin of this concept
lies in attempts to fill the blanks in the Everett's relative-state
interpretation, and to make it a fully valid description of physical
reality. To achieve this, it was necessary to consider not only the
main system interacting with the measurement apparatus (like von Neumann
and Everett did) but also the role of the environment in eliminating
correlations between different possible measurements when interacting
with the measurement apparatus. The interaction of the environment
with the main system (and the measurement apparatus) is the core of
the decoherence theory, which followed Everett's thesis. In this article,
we review the measurement process according to von Neumann, Everett's
relative state interpretation, the purpose of decoherence and some
of its follow-up until Wojciech Zurek's primordial paper that consolidated
the concept of pointer state, previously presented by Heinz Dieter
Zeh. Employing a simple physical model consisting of a pair of two-level
systems -- one representing the main system, the other the measurement
apparatus -- and a thermal bath -- representing the environment --
we show how pointer states emerge, explaining its contributions to
the question of measurement in quantum mechanics, as well as its limitations.
Finally, we briefly show some of its consequences. This paper is accessible
to readers with elementary knowledge about quantum mechanics, on the
level of graduate courses. 
\end{abstract}

\section{Introduction}

\textcolor{black}{Most lay people who come in contact with quantum
mechanics learn about it through some paradox, usually Schrödinger's
cat}\footnote{\textcolor{black}{See the English translation of Schrödinger's original
paper at \cite{key-1}.}}\textcolor{black}{. The counter-intuitive idea that a cat can simultaneously
be alive and dead inside a box before it is observed is puzzling and
highlights the troubling aspects of quantum theory. But lay people
are not the only ones who are confused by it.}

\textcolor{black}{Since its inception, the question of interpretation
of quantum mechanics has eluded physicists \cite{key-2,key-3,key-4}
and early on its concepts have been the subject of heated debate \cite{key-2}.
In quantum mechanics, physical objects are described by }\textcolor{black}{\emph{wave
functions}}\textcolor{black}{{} or }\textcolor{black}{\emph{state vectors}}\textcolor{black}{{}
\cite{key-5}, mathematical structures that evolve unitarily while
they are not being measured by a macroscopic apparatus, in which case
they acquire a single value that in general can only be predicted
probabilistically. Such state vectors can be written as a superposition
of (eigen)states corresponding to the possible (eigen)values of a
measurement, in a linear combination where the coefficients are related
to the probabilities for each measurement result. All these alternatives
for measurement results not only may be considered for calculations,
but can also interfere with each other. This is the so-called }\textcolor{black}{\emph{superposition
principle }}\textcolor{black}{- a physical description of which can
be found in \cite{key-2,key-60}.}

\textcolor{black}{The idea that objects such as particles could be
described as non-localized waves was already challenging enough, but
relinquishing determinism at least in theory proved to be too much
to many classical physicists. Among the several paths to understand
the emergence of the classical world, Everett's relative-state interpretation
\cite{key-17,key-18,key-19,key-20,key-21,key-24} originated the decoherence
theory \cite{key-4,key-39,key-44,key-45,key-46,key-47,key-61,key-62,key-63,key-64,key-65,key-66}
- the idea that it is the influence of the environment on the principal
system and the measurement apparatus that causes the emergence of
the classical world.}

\textcolor{black}{Research for this purpose was being conducted \cite{key-7,key-31,key-67},
when an article \cite{key-6} treating a question of interpretation
that had been mostly ignored by mainstream physics for decades unleashed
a revolution that affected both philosophical and practical aspects
of quantum mechanics by exploring the incipient concept of pointer
state \cite{key-7}. This idea was introduced to help to solve one
of the difficulties created by the superposition principle, but it
had deeper consequences in many correlated areas of quantum physics
\cite{key-39,key-44,key-45,key-46,key-47,key-68}.}

\textcolor{black}{Although there are several papers about the nature
of the pointer states, since the pioneer works of Zeh \cite{key-7,key-31}
and Zurek \cite{key-6}, with several reviews of the latter \cite{key-4,key-44,key-46,key-77}
- and a very good historical review in \cite{key-39} - these papers
are, in general, or too technical or too philosophical, frequently
treating more than the specific concept of pointer states and with
no concessions to the undergraduate student. To fill this gap, here
we deal mainly with the importance of pointer states and the influence
it had in quantum mechanics. We illustrate our approach with a simple
but realistic example of emergence of a pointer basis that can be
understood with a graduate-level knowledge of quantum mechanics \cite{key-8}.}

\textcolor{black}{But, to explain it, we shall first review the historical
progress that led to the difficulties it attempted to solve, and then
proceed to explain how it is formulated. In Sec. 2, we provide an
overview of the history to show operationally what is a relative state
and how it is linked to such concepts as post-selection. In Sec. 3,
we explain how measurement can be pictured in a realist interpretation
such as Dirac-von Neumann's or Everett-based interpretations. In Sec.
4, we make a simple calculation of how the pointer basis is obtained,
illustrating step-by-step the whole process of interactions between
the principal system and the measurement apparatus, and between the
measurement apparatus and the environment. The calculations may seem
complicated at the first sight, but they are explained in detail in
the appendices and demand from the reader just the fundamental knowledge
of the harmonic oscillator operators. We see a brief overview of further
developments of the theory in Sec. 5 and conclude in Sec. 6.}

\section{The development of quantum interpretations}

The commonplace realist interpretation of classical physics, in which
every mathematical concept of the theory was thought to correspond
to an element of reality was soon replaced by a near consensus around
the ``orthodox'' Copenhagen interpretation \cite{key-2,key-3,key-9,key-10},
championed by Niels Bohr and centered around the Institute for Theoretical
Physics of the University of Copenhagen. In this view, the wave function
did not correspond to any real object, but simply encapsulated all
the knowledge we had about the quantum system at a given time. The
Schrödinger's cat problem \cite{key-1} is an emblematic example:
the discontinuous process in which ``opening the box'' revealed
the ``cat'' to be ``dead'' or ``alive'' does not correspond
to any change in the world, but to a process of acquisition of information
(it is important to note, however, that the ``cat'' is not dead
or alive before the box is open, but it is both dead \emph{and} alive,
i.e., the cat state is a superposition of both alternatives, an aspect
of vital importance - see sec. 3.2). Questions about the nature of
the quantum world or what each mathematical step symbolized were irrelevant,
and unanswerable: the important thing was that the theory predicted
correctly what the macroscopic apparatus -- this one indeed real and
tenable -- was capable of detecting. Such pragmatism proved especially
useful in theoretical and experimental research programs where only
the predictions of the measurement were necessary, and philosophical
discussions of their meaning could be relinquished to a secondary
level. 

Concomitantly to this, a more informal interpretation of quantum mechanics
emerged in the 1920s, together with the first textbooks on the subject
\cite{key-5,key-60}, and for this reason it is sometimes called the
\emph{von Neumann-Dirac }interpretation (due to P. A. M. Dirac's pioneer
work \cite{key-11}) - also referred to as ``Princeton school''
\cite{key-12}, albeit more rarely so. These scientists treated wave
functions as more or less real objects, which were nevertheless subject
to two different kinds of evolution: the random \emph{collapse} due
to measurements; and the continuous unitary evolution described by
the Schrödinger equation. von Neumann named these \emph{Process 1}
and \emph{Process 2} \cite{key-5}, a notation that was reproduced
in some later works - including Everett's \cite{key-17,key-18,key-24}.

The asymmetry between these processes was intriguing, especially after
von Neumann proved \cite{key-5} that the stochastic \emph{Process
1} could not be explained simply as a special case of the deterministic
\emph{Process 2,} thus making clear its axiomatic nature. There seemed
to be an unbreakable frontier between the micro and macro worlds,
suggesting that quantum mechanics could not be a universal theory,
that it had to be at the very least patched or altogether replaced
by something entirely new.

Some physicists, seeing an opportunity to recover determinism, came
up with the idea that maybe the wave function description of reality
did not cover the whole picture, and that maybe some extra elements
were necessary to fully explain the workings of the universe. These
approaches became known as \emph{hidden variables }interpretations
\cite{key-3}. Bohm became the name best associated with this program
due to his formalization of the theory in the 1950s \cite{key-13,key-14,key-15},
but many physicists had suggested similar approaches before. Among
these we can count Einstein, who highlighted in the famous \emph{EPR
}article \cite{key-16} that superposition was possible between the
product state of objects distant from each other, so that a measurement
in one could cause a collapse far away. The situation where the state
of a system became dependent of the state of another through the superposition
principle was later called \emph{entanglement} by Schrödinger \cite{key-69}.
The entanglement is one of the most notable aspects of the quantum
theory, originating a true field of investigations on technologies
such as quantum information \cite{key-39,key-36,key-35,key-70}.

In the end of the 1950s, Hugh Everett III \cite{key-17,key-18} proposed
a distinct realistic approach that held that quantum mechanics was
thoroughly correct, explaining away \emph{Process 1} as merely the
subjective effect of entangling the observer to the system under observation.
From the point of view of the entangled observer, the quantum system
would \emph{appear} to have acquired a determined \emph{relative state},
but the universal wave function would still be a superposition state
evolving unitarily according to the Schrödinger equation. We will
see this in more detail below.

\subsection{Relative states\label{sec:A-S}}

The concept of relative state, which naturally arose from the mathematical
formalization of quantum mechanics \cite{key-5}, is instrumental
in understanding the pointer basis problem, so we will briefly illustrate
it here. As its name suggests, a relative state can only be determined
in respect to the state of another system with which it is interacting.
For this reason, suppose we have two interacting systems, $S$ and
$A$, with corresponding Hilbert spaces $\mathcal{H}_{S}$ and $\mathcal{H}_{A}$
and bases $\left\{ \left|s_{i}\right\rangle \right\} $ and $\left\{ \left|a_{i}\right\rangle \right\} $.
Any normalized state of the joint system $S+A$ can be written as:

\begin{equation}
\left|\psi\right\rangle _{S+A}=\underset{i,j}{\sum}c_{ij}\left|s_{i}\right\rangle _{S}\left|a_{j}\right\rangle _{A},
\end{equation}
where $\sum_{i,j}\left|c_{ij}\right|^{2}=1$, and with a tensor product
implied between $\left|s_{i}\right\rangle _{S}$ and $\left|a_{j}\right\rangle _{A}$,
that is, $\left|s_{i}\right\rangle _{S}\left|a_{j}\right\rangle _{A}=\left|s_{i}\right\rangle _{S}\otimes\left|a_{j}\right\rangle _{A}$
represents a state in which both systems are in well-determined states
of their respective bases. A state written simply like $\left|s\right\rangle _{S}\otimes\left|a\right\rangle _{A}$
is \emph{separable}, we can divide it into a unique state vector for
each system.

However, many times when a global state vector like $\left|\psi\right\rangle _{S+A}$
is defined, subsystems $S$ and $A$ are no longer independent of
each other. This means that we cannot split the system in two with
well-defined substates unless we choose what is the state of one of
them. The precise state vector of $S$, then, is conditioned on the
state of $A$, which is precisely the nature of entanglement \cite{key-16,key-69,key-70}.
The motivation behind the concept of relative state was the need to
express this conditional state.

Therefore, if the state of system $A$ is chosen to be $\left|a_{k}\right\rangle _{A}$,
the relative state of system $S$ will be

\begin{equation}
\left|\psi_{R}\left(a_{k}\right)\right\rangle _{S}=N_{k}\underset{i}{\sum}c_{ik}\left|s_{i}\right\rangle _{S}
\end{equation}
where $N_{k}$ is a normalization constant given by:

\begin{equation}
N_{k}=\frac{1}{\sum_{i}\left|c_{ik}\right|^{2}}=\frac{1}{\left|\left\langle a_{k}\right.\left|\psi\right\rangle _{S+A}\right|^{2}}.
\end{equation}

This procedure is operationally equivalent to imposing a final state,
that is, to post-selection.
\[
\left|\psi_{R}\left(a_{k}\right)\right\rangle _{S}=\frac{\left\langle a_{k}\right.\left|\psi\right\rangle _{S+A}}{\left|\left\langle a_{k}\right.\left|\psi\right\rangle _{S+A}\right|^{2}}.
\]
In other words, we are saying that, given that we know what is the
state of $A$ (through a measurement, for example), we have determined
the state of $S$. As it is, this relative state furnishes the probability
distribution of measurements on $S$ once a measurement on $A$ is
known to have resulted in $a_{k}$. Or, if $A$ is the observer, we
can say that if it believes it is on the state $\left|a_{k}\right\rangle $
(as a result of the knowledge acquired during the measurement, for
example), then it will have the \emph{subjective} impression that
the wavefunction of $S$ has collapsed into the appropriate relative
state.

Everett intended, by these means, to explain the measurement process
or, more precisely, the wave function reduction. The apparatus would
perceive a single definite value of a system property but the global
wave function would remain a superposition of terms where the system
has a specific value for a property of interest and the apparatus
had its state altered to reflect that property.

Each measurement/interaction ``branches'' the state of the observer
into different eigenstates. Each branch represents a different result
of the measurement, which corresponds to an eigenstate for the main
system $S$. All the branches coexist in the superposition \emph{after}
the measurement (or series of measurements). According to Everett,
this transition of the state superposition to the single value found
in the measurement does not exist and is not necessary to conciliate
his theory with the universe. To him, all branches are equally real.

Such a description, while having the advantage of getting rid of the
asymmetry between the two evolution processes and re-affirming the
universality of quantum mechanics, still required an additional interpretation
for the meaning of leaving the observer in a superposition state.
Hugh Everett did not spell it out in his original paper \cite{key-18},
but attributing the same level of reality for each and every of these
\emph{branches} was soon understood to mean that the universe itself
was being fractured in many a different worlds, each of which differed
in the result observed, but each equally unaware of the existence
of the others.

Everett's work is centered on the concept of relative state \cite{key-20,key-21}
and, although it never gathered as many followers as ``orthodox''
interpretations, it is taken seriously by many \cite{key-22,key-23}.
One of its developments was the Many-Worlds Interpretation (MWI) \cite{key-24},
as Bryce DeWitt named it to stimulate debate, was soon to raise objection,
some of a more aesthetic order: it was inelegant, absurd and unnecessary
to fill the universe endless invisible worlds \cite{key-19,key-25,key-26,key-27,key-28,key-29}.
Others, however, identified more tangible shortcomings in the theory,
which could be broadly classified in two kinds \cite{key-27}:
\begin{enumerate}
\item \textbf{How to conciliate probabilities with multiple universes.}
If all possible events do occur, what is the meaning of a number such
as probability? How does the Born rule emerge from the coefficients
of superpositions? That is, why do we have the sensation that a two-level
state collapses to certain value with probability that is the square
norm of one of the coefficients rather than, say, $1/2$? \cite{key-30}
\item \textbf{Why do different branches of the universe fail to interfere
with each other?} As Zeh noted in the 1970s, the superposition state
system-object + observer keeps evolving in a unitary manner, according
to the Schrödinger equation \cite{key-31}. Why does the universe
behave in such a way as if the the other branches did not exist?
\end{enumerate}
An off-shoot of the second kind of problem was also highlighted in
the 1970s \cite{key-31}. According to experience, repeated measurements
by the same (well-designed) macroscopic apparatus result all in end
states in the same basis, but in the Everettian description we could
in principle have an infinity of possible basis to have the measurement
performed at. How is the effective basis we observe selected? How
to eliminate this basis ambiguity?

\subsection{Basis ambiguity\label{sub:Basis-ambiguity}}

To further understand the problem of basis ambiguity we must remember
that in quantum mechanics different measurements can interfere with
each other. This happens when the observables referring to these measurements
do not commute \footnote{in other words, quantum mechanics is a \emph{non-contextual theory}
\textcolor{blue}{\cite{key-59}}, as opposed to classical mechanics,
but a more in-depth discussion of the concept of \emph{contextuality}
is beyond the scope of this article}. This fact is directly related to \emph{Heisenberg's uncertainty
principle} \cite{key-1}, which states limits within which one can
determine the value of non-commuting observables.

A measurement associated with a given observable demands an experimental
setup and an interaction of the measurement apparatus with the physical
system from which we intend to extract information. Suppose we want
to measure one of two given quantities associated to the observables
$\hat{o}_{1}$ and $\hat{o}_{2}$ (they can refer to position, momentum,
spin component, energy, etc.) and, in a given instant, we perform
the measurement of $\hat{o}_{1}$ - which evidently interfered in
the state of the system. To measure $\hat{o}_{2}$ right in sequence,
we must analyse the commutator of both observables, $\left[\hat{o}_{1},\hat{o}_{2}\right]$,
\begin{itemize}
\item $\left[\hat{o}_{1},\hat{o}_{2}\right]=0$: it is possible to measure
$\hat{o}_{2}$ employing the same experimental setup used to measure
$\hat{o}_{1}$ and the order of measurements will not affect the results;
\item $\left[\hat{o}_{1},\hat{o}_{2}\right]\neq0$: the measurement of $\hat{o}_{2}$
requires a different, incompatible experimental setup, and its results
will be influenced by the prior measurement of $\hat{o}_{1}$.
\end{itemize}
Let us in this example restrict our discussion to the observables

\begin{equation}
\begin{cases}
\hat{o}_{1} & =\hat{\sigma}_{z}\\
\hat{o}_{2} & =x\hat{\sigma}_{x}+y\hat{\sigma}_{y}
\end{cases},\:x\pm iy\neq0
\end{equation}
where $\hat{\sigma}_{x},\:\hat{\sigma}_{y}$ and $\hat{\sigma}_{z}$
are the Pauli matrices \cite{key-8},

\begin{equation}
\hat{\sigma}_{x}=\left(\begin{array}{cc}
0 & 1\\
1 & 0
\end{array}\right),\:\hat{\sigma}_{y}=\left(\begin{array}{cc}
0 & -i\\
i & 0
\end{array}\right),\:\hat{\sigma}_{z}=\left(\begin{array}{cc}
1 & 0\\
0 & -1
\end{array}\right)
\end{equation}
on the basis of eigenstates of $\hat{\sigma}_{z},$$\left\{ \left|+\right\rangle ,\left|-\right\rangle \right\} $
- evidently $\left[\hat{o}_{1},\hat{o}_{2}\right]\neq0$. The matrix
representation of $\hat{o}_{2}$ will be

\begin{equation}
\hat{o}_{2}=\left(\begin{array}{cc}
0 & x-iy\\
x+iy & 0
\end{array}\right).
\end{equation}
Its eigenvalues are $\pm\sqrt{x^{2}+y^{2}}$ and we label the corresponding
eigenstates with $\left|\pm\sqrt{x^{2}+y^{2}}\right\rangle \equiv\left|\pm\right\rangle _{xy}$
so that

\begin{equation}
\begin{cases}
\left|+\right\rangle _{xy} & =\sqrt{\frac{1}{2}\frac{x-iy}{x+iy}}\left|+\right\rangle +\frac{1}{\sqrt{2}}\left|-\right\rangle \\
\left|-\right\rangle _{xy} & =\sqrt{\frac{1}{2}\frac{x-iy}{x+iy}}\left|+\right\rangle -\frac{1}{\sqrt{2}}\left|-\right\rangle 
\end{cases}\label{eq:xyEigenstates}
\end{equation}
where the $\left|\pm\right\rangle $ are the eigenstates of $\hat{o}_{1}=\hat{\sigma}_{z}$.
These can be written in terms of the eigenstates from Eq. (\ref{eq:xyEigenstates})
if we perform an inversion,

\begin{equation}
\begin{cases}
\left|+\right\rangle  & =\sqrt{\frac{1}{2}\frac{x+iy}{x-iy}}\left(\left|+\right\rangle _{xy}+\left|-\right\rangle _{xy}\right)\\
\left|-\right\rangle  & =\frac{\left|+\right\rangle _{xy}-\left|-\right\rangle _{xy}}{\sqrt{2}}
\end{cases}.
\end{equation}

To describe the apparatus performing the measurement, we will employ
another quantum system that ends correlated with the observed state,
an approach that has been applied elsewhere \cite{key-5,key-18,key-7}.
In the end, the eigenstates (whose associated eigenvalues are the
possible results of our measurement) should be correlated with distinguishable,
stable states of the observer, called at times ``memory states''
\cite{key-31}. To simplify, here we will consider the observer as
well as a two-level system.

For example, the joint state of the system and the observer after
a pre-measurement of $\hat{o}_{1}$ - that is, after the interaction
with the observer is complete, but before it has suffered a wave function
collapse - is:

\begin{equation}
\left|\psi\right\rangle =a\left|++\right\rangle +b\left|--\right\rangle .\label{psiz}
\end{equation}
(remembering that $\left|\pm\pm\right\rangle \equiv\left|\pm\right\rangle \left|\pm\right\rangle $).
Implicitly, here, we are treating the process of information acquisition
as involving two steps:
\begin{itemize}
\item Pre-measurement, where the interaction between system and apparatus
creates correlations between the two;
\item The random collapse, defining the possible results of the measurement.
\end{itemize}
Re-writing this result in terms of the eigenbasis $\left\{ \left|++\right\rangle _{xy},\left|+-\right\rangle _{xy},\left|-+\right\rangle _{xy},\left|--\right\rangle _{xy}\right\} $
of $\hat{o}_{2}$ (where $\left|\pm\pm\right\rangle _{xy}\equiv\left|\pm\right\rangle _{xy}\left|\pm\right\rangle _{xy}$),
we find

\begin{eqnarray}
\left|\psi\right\rangle  & = & \frac{1}{2}\left(a\frac{x+iy}{x-iy}+b\right)\left(\left|++\right\rangle _{xy}+\left|--\right\rangle _{xy}\right)\nonumber \\
 & + & \frac{1}{2}\left(a\frac{x+iy}{x-iy}-b\right)\left(\left|+-\right\rangle _{xy}+\left|-+\right\rangle _{xy}\right).\label{psixy}
\end{eqnarray}
If in (\ref{psixy}) we choose $b=a\frac{x+iy}{x-iy}$ (or $b=-a\frac{x+iy}{x-iy}$),
we will have found the expected state after a pre-measurement of the
observable $\hat{o}_{2}$. However, by definition $\left[\hat{o}_{2},\hat{o}_{1}\right]\neq0$,
so it should not be possible to measure both observables with the
same experimental setup. 

This is an example of the problem of basis ambiguity, which cannot
be solved through unitary operations and state vectors. Unless we
impose a collapse, we cannot be sure which basis is being measured.
An Everettian interpretation would require an additional criterion
for deciding the basis of the measurer's pointer \cite{key-31}. The
problem is not restricted to an interpretation without state vector
reduction, however. Even more mundane interpretations of quantum mechanics
would require an explanation as to why only position and momentum
of a macroscopic object are determined rather than any other basis
of it \cite{key-32}.

As will become clearer later, the solution for this problem involves
considering the effect of an external environment on the process of
performing the measurement. To show this, however, we will require
a model for the dynamics of quantum measurements.

\section{Quantum measurements according to von Neumann\label{sec:pre-mes}}

A simple model for quantum measurements, employed multiple times -
including by Everett - to describe Stern-Gerlach-type apparatuses
\cite{key-8}, was proposed in von Neumann's classical textbook \cite{key-5},
where it is described as a correlation between the system-object and
the observer. For example,

\begin{equation}
\left|\psi\left(0\right)\right\rangle =\left(a\left|+\right\rangle +b\left|-\right\rangle \right)\left|+\right\rangle \rightarrow\left|\psi\left(t\right)\right\rangle =a\left|+\right\rangle \left|+\right\rangle +b\left|-\right\rangle \left|-\right\rangle .\label{medidavN}
\end{equation}
A careful reader will notice that this kind of evolution produces
exactly the kind of ``pre-measurement'' state mentioned in the previous
section.

Now, if we describe the evolution by $\left|\psi\left(t\right)\right\rangle =\hat{\Delta}\left|\psi\left(0\right)\right\rangle $,
what should be the form of the time evolution operator $\hat{\Delta}$?
In the basis $\left\{ \left|++\right\rangle ,\left|+-\right\rangle ,\left|-+\right\rangle ,\left|--\right\rangle \right\} $,
we can write the matrix that leads to that final state as

\begin{equation}
\left(\begin{array}{cccc}
\Delta_{11} & \Delta_{12} & \Delta_{13} & \Delta_{14}\\
\Delta_{21} & \Delta_{22} & \Delta_{23} & \Delta_{24}\\
\Delta_{31} & \Delta_{32} & \Delta_{33} & \Delta_{34}\\
\Delta_{41} & \Delta_{42} & \Delta_{43} & \Delta_{44}
\end{array}\right)\left(\begin{array}{c}
a\\
0\\
b\\
0
\end{array}\right)=\left(\begin{array}{c}
a\\
0\\
0\\
b
\end{array}\right),
\end{equation}
which corresponds to the system of equations

\begin{equation}
\begin{cases}
a\Delta_{11}+b\Delta_{13}= & a\\
a\Delta_{21}+b\Delta_{23}= & 0\\
a\Delta_{31}+b\Delta_{33}= & 0\\
a\Delta_{41}+b\Delta_{43}= & b
\end{cases}.
\end{equation}

If this system of equations must be satisfied for any value $a$ and
$b$, we shall have an evolution operator $\hat{\Delta}$ of the type:

\[
\hat{\Delta}=\left(\begin{array}{cccc}
1 & \Delta_{12} & 0 & \Delta_{14}\\
0 & \Delta_{22} & 0 & \Delta_{24}\\
0 & \Delta_{32} & 0 & \Delta_{34}\\
0 & \Delta_{42} & 1 & \Delta_{44}
\end{array}\right).
\]

The remaining elements $\Delta_{12},\Delta_{22},\Delta_{32},\Delta_{34}$
and $\Delta_{14},\Delta_{24},\Delta_{34},\Delta_{44}$ may not be
freely chosen because a quantum evolution operator is subject to the
additional unitarity condition. That is,

\begin{equation}
\hat{\Delta}\left(t\right)\hat{\Delta}^{\dagger}\left(t\right)=\mathbf{1}_{4\times4},
\end{equation}
which translates into the following system of equations:

\begin{equation}
\begin{cases}
\left|\Delta_{12}\right|^{2}+\left|\Delta_{14}\right|^{2}= & 0\\
\left|\Delta_{42}\right|^{2}+\left|\Delta_{44}\right|^{2}= & 0\\
\left|\Delta_{22}\right|^{2}+\left|\Delta_{24}\right|^{2}= & 1\\
\left|\Delta_{32}\right|^{2}+\left|\Delta_{34}\right|^{2}= & 1\\
\Delta_{22}\Delta_{32}^{*}+\Delta_{24}\Delta_{34}^{*}= & 0
\end{cases}.
\end{equation}

From the two first equations we conclude that $\Delta_{12}=\Delta_{14}=\Delta_{42}=\Delta_{44}=0$,
because the sum of positive numbers is only zero when all the terms
are zero. From the other three, we write the remaining four unknown
elements as:

\begin{equation}
\begin{cases}
\Delta_{22}= & e^{i\phi_{22}}\cos\theta\\
\Delta_{24}= & e^{i\phi_{24}}\sin\theta\\
\Delta_{32}= & e^{i\phi_{32}}\cos\vartheta\\
\Delta_{34}= & -e^{i\left(\phi_{24}+\phi_{32}-\phi_{22}\right)}\sin\vartheta
\end{cases},\label{solsDeltas}
\end{equation}
where we restrict the angles $\theta,\vartheta$ to the first quadrant,
so that all information about signs is included in the complex phases
$\phi_{22},\phi_{24},\phi_{32}$.

The last equation allows us to conclude that $\cos\left(\theta+\vartheta\right)=0$,
which in the first quadrant means that $\vartheta=\pi/2-\theta$.
Therefore, the most general form of the time evolution operator is

\begin{equation}
\hat{\Delta}=\left(\begin{array}{cccc}
1 & 0 & 0 & 0\\
0 & \Delta_{22} & 0 & e^{i\phi_{24}}\sqrt{1-\left|\Delta_{22}\right|^{2}}\\
0 & e^{i\phi_{32}}\sqrt{1-\left|\Delta_{22}\right|^{2}} & 0 & -e^{i\left(\phi_{24}+\phi_{32}\right)}\Delta_{22}^{*}\\
0 & 0 & 1 & 0
\end{array}\right).\label{DELTAgeral}
\end{equation}

The choice of the three phases is arbitrary, but the modulus of $\Delta_{22}$
must be kept smaller or equal to one, because it is limited by the
value of cosine, according to Eq. (\ref{solsDeltas}).

This is just an evolution operator for a static instant, however.
To find the complete dynamics of the system, we need a candidate Hamiltonian
capable of generating this evolution.

\subsection{Candidate object-apparatus interaction}

A possible object-apparatus interaction that produces$\hat{\Delta}$
is \cite{key-4}:

\begin{eqnarray}
\hat{H}_{SA} & = & g\left(\left|-\right\rangle \left\langle -\right|\right)_{S}\left[\left(\left|-\right\rangle \left\langle -\right|\right)_{x}\right]_{A}\nonumber \\
 & = & \frac{g}{4}\left(\hat{1}-\hat{\sigma}_{z}\right)_{S}\left(\hat{1}-\hat{\sigma}_{x}\right)_{A}\label{HsaZurek}
\end{eqnarray}

To verify this, let us obtain its matrix form in the basis $\left\{ \left|++\right\rangle ,\left|+-\right\rangle ,\left|-+\right\rangle ,\left|--\right\rangle \right\} $:

\begin{equation}
\hat{H}_{SA}=\frac{g}{4}\left(\begin{array}{cc}
0 & 0\\
0 & 2
\end{array}\right)\otimes\left(\begin{array}{cc}
1 & -1\\
-1 & 1
\end{array}\right)=\frac{g}{2}\left(\begin{array}{cccc}
0 & 0 & 0 & 0\\
0 & 0 & 0 & 0\\
0 & 0 & 1 & -1\\
0 & 0 & -1 & 1
\end{array}\right)
\end{equation}
The eigenvalues of this Hamiltonian will be a triply-degenerate $0$
and a $g$. Choosing three orthonormal vectors within the subspace
of eigenvalue zero, we find a possible eigenbasis for the matrix:

\begin{eqnarray}
\left|0\right\rangle _{1} & = & \left|++\right\rangle \nonumber \\
\left|0\right\rangle _{2} & = & \left|+-\right\rangle \nonumber \\
\left|0\right\rangle _{3} & = & \frac{\left|-+\right\rangle +\left|--\right\rangle }{\sqrt{2}}\nonumber \\
\left|g\right\rangle  & = & \frac{\left|-+\right\rangle -\left|--\right\rangle }{\sqrt{2}}.
\end{eqnarray}
Regardless of the precise choice, the Hamiltonian in the diagonal
basis will be:

\begin{equation}
\hat{H}_{SA}=\left(\begin{array}{cccc}
0 & 0 & 0 & 0\\
0 & 0 & 0 & 0\\
0 & 0 & 0 & 0\\
0 & 0 & 0 & g
\end{array}\right).
\end{equation}
The diagonal time-evolution operator can be found simply by taking
the exponential of the diagonal elements:

\begin{equation}
\hat{u}\left(t\right)=e^{-i\frac{t}{\hbar}\hat{H}_{SA}}=\left(\begin{array}{cccc}
1 & 0 & 0 & 0\\
0 & 1 & 0 & 0\\
0 & 0 & 1 & 0\\
0 & 0 & 0 & e^{-i\frac{gt}{\hbar}}
\end{array}\right).
\end{equation}
We return to the representation in the canonical basis $\left\{ \left|++\right\rangle ,\left|+-\right\rangle ,\left|-+\right\rangle ,\left|--\right\rangle \right\} $
by employing the eigenvector matrix

\begin{equation}
\hat{M}=\left(\begin{array}{cccc}
1 & 0 & 0 & 0\\
0 & 1 & 0 & 0\\
0 & 0 & \frac{1}{\sqrt{2}} & \frac{1}{\sqrt{2}}\\
0 & 0 & \frac{1}{\sqrt{2}} & -\frac{1}{\sqrt{2}}
\end{array}\right)=\hat{M}^{-1}.
\end{equation}

Hence, the time evolution operator in the initial basis will be

\begin{equation}
\hat{U}\left(t\right)=\hat{M}\hat{u}\left(t\right)\hat{M}^{-1}=\left(\begin{array}{cccc}
1 & 0 & 0 & 0\\
0 & 1 & 0 & 0\\
0 & 0 & \frac{1+e^{-i\frac{g}{\hbar}t}}{2} & \frac{1-e^{i\frac{g}{\hbar}t}}{2}\\
0 & 0 & \frac{1-e^{-i\frac{g}{\hbar}t}}{2} & \frac{1+e^{-i\frac{g}{\hbar}t}}{2}
\end{array}\right).\label{Uz}
\end{equation}

Comparing (\ref{Uz}) and (\ref{DELTAgeral}), we see that if, in
(\ref{Uz}), we choose the time $\tau_{\mathrm{pm}}$ such that

\begin{equation}
\tau_{\mathrm{pm}}=n\frac{\pi\hbar}{g},\:n\text{ odd}\label{tpm}
\end{equation}
we find

\begin{equation}
\hat{U}\left(\tau_{\mathrm{pm}}\right)=\left(\begin{array}{cccc}
1 & 0 & 0 & 0\\
0 & 1 & 0 & 0\\
0 & 0 & 0 & 1\\
0 & 0 & 1 & 0
\end{array}\right).\label{Uztau}
\end{equation}

This is a valid operator $\hat{\Delta}$, as it is equivalent to (\ref{DELTAgeral})
when you choose $\Delta_{22}=1$ and take the phases $\phi_{24}$
and $\phi_{32}$ so that their complex exponential cancels out the
minus sign in front of the $\Delta_{34}$ matrix element.

Now, suppose we start with the observed system in a superposition
state $a\left|+\right\rangle +b\left|-\right\rangle $, while the
observer is initially in the state $\left|+\right\rangle $. We make
this choice to highlight that the system we want to measure can be
in any superposition state, but the observer is in a fixed, known
one. After any given time length $t$, the evolution of $\left|\psi\left(0\right)\right\rangle =a\left|++\right\rangle +b\left|-+\right\rangle $
under the Hamiltonian (\ref{HsaZurek}) will result in the state

\begin{equation}
\left|\psi\left(t\right)\right\rangle =a\left|++\right\rangle +\frac{1+e^{-i\frac{g}{\hbar}t}}{2}b\left|-+\right\rangle +\frac{1-e^{-i\frac{g}{\hbar}t}}{2}b\left|--\right\rangle ,
\end{equation}
which simplifies, at $t=\tau_{\mathrm{pm}}$, to the entangled state
we were expecting

\begin{equation}
\left|\psi\left(\tau_{\mathrm{pm}}\right)\right\rangle =a\left|++\right\rangle +b\left|--\right\rangle ,\label{psitm}
\end{equation}
which is what we had called the state of pre-measurement of system
$S$ in Sec. I. In this case, we have found the entangled joint state
of the system and observer just before the measurement of the observable
$\hat{\sigma}_{z}$. However, this state does not include the final
wave function collapse that we observe in a quantum system (or, if
thought within an Everettian framework, is still subject to the basis
ambiguities mentioned above). To describe the stochastic \emph{process
2}, the state vector is not sufficient, for we require an ensemble
description of the quantum system \cite{key-31}.

\subsection{Ensembles and probabilities\label{sub:probabilities}}

The density operator \textcolor{blue}{-} introduced by von Neumann
in his book \cite{key-5}, Furry on \cite{key-71} and by L. D. Landau
in his paper published in Z. Phys. 1927 (translation available in
\cite{key-34}) - is an especially useful tool in the treatment of
open systems, the evolution of which is not as simple and unitary
as the one provided by the Schrödinger equation \cite{key-8,key-35,key-36}.
It is, therefore, useful in the description of a inherently irreversible
evolution such as the one of the wave function collapse.

The reader may already familiar with this formalism to the level discussed
in\cite{key-8}, but we shall make a brief review. Given a quantum
state vector $\left|\psi\left(t\right)\right\rangle $, its associated
density vector $\hat{\rho}\left(t\right)$ is defined as

\begin{equation}
\hat{\rho}\left(t\right)=\left|\psi\left(t\right)\right\rangle \left\langle \psi\left(t\right)\right|.\label{puro}
\end{equation}
Considering that the state $\left|\psi\left(t\right)\right\rangle $
evolved from an initial state $\left|\psi\left(0\right)\right\rangle $
through a unitary operator $\hat{U}\left(t\right)$, that is,

\begin{equation}
\left|\psi\left(t\right)\right\rangle =\hat{U}\left(t\right)\left|\psi\left(0\right)\right\rangle ,
\end{equation}
the evolved density operator will be simply

\begin{equation}
\hat{\rho}\left(t\right)=\hat{U}\left(t\right)\left|\psi\left(0\right)\right\rangle \left\langle \psi\left(0\right)\right|\hat{U}^{\dagger}\left(t\right).
\end{equation}

The time evolution of the density operator of a closed system is given,
therefore, by the Liouville-von Neumann equation \cite{key-5}, derived
from Schrödinger's equation applied both to the state vector and its
dual:

\begin{equation}
\frac{d}{dt}\hat{\rho}\left(t\right)=-\frac{i}{\hbar}\left[\hat{H},\hat{\rho}\right].\label{eqLvN}
\end{equation}

The greatest advantage of the density operator is the possibility
of describing a system, the precise state $\left|\psi\left(t\right)\right\rangle $
of which is unknown, but to which we can attribute probabilities $p_{1}$,
$p_{2}$,$p_{3}$,...,$p_{n}$ of finding it in any of the states
$\left|\psi_{1}\left(t\right)\right\rangle $,\emph{ }$\left|\psi_{2}\left(t\right)\right\rangle $,
$\left|\psi_{3}\left(t\right)\right\rangle $, \emph{...,} $\left|\psi_{n}\left(t\right)\right\rangle $.
In this case, the mixture of possible states is represented by the
density operator:

\begin{eqnarray}
\hat{\rho}\left(t\right) & = & p_{1}\left|\psi_{1}\left(t\right)\right\rangle \left\langle \psi_{1}\left(t\right)\right|+p_{2}\left|\psi_{2}\left(t\right)\right\rangle \left\langle \psi_{2}\left(t\right)\right|+...+p_{n}\left|\psi_{n}\left(t\right)\right\rangle \left\langle \psi_{n}\left(t\right)\right|\nonumber \\
 & = & \overset{n}{\underset{j=1}{\sum}}p_{j}\left|\psi_{j}\left(t\right)\right\rangle \left\langle \psi_{j}\left(t\right)\right|.\label{mistura}
\end{eqnarray}
It is important to emphasize that (\ref{mistura}) does not represent
a superposition of states $\left|\psi_{j}\left(t\right)\right\rangle $.
The coefficients $p_{j}$ are all real \emph{classical probabilities}
that satisfy $\overset{n}{\underset{j=1}{\sum}}p_{j}=1$, and do not
interfere quantically. The density operator (\ref{mistura}) represents
a \emph{statistical ensemble} or \emph{mixture}, while (\ref{puro})
is a \emph{pure state}. Both obey the Liouville-von Neumann equation
(\ref{eqLvN}), due to the linearity of the construction of the mixture
operator.

To illustrate the difference between a pure state and a classical
mixture, let us consider a simple two-state, with basis $\left\{ \left|+\right\rangle ,\left|-\right\rangle \right\} $,
and two possible states

\begin{equation}
\begin{cases}
\left|\psi_{1}\right\rangle  & =a_{1}\left|+\right\rangle +a_{2}\left|-\right\rangle \\
\left|\psi_{2}\right\rangle  & =b_{1}\left|+\right\rangle +b_{2}\left|-\right\rangle 
\end{cases}
\end{equation}
(this is a simple set-up where we are not dealing with pre-measurements,
distinctions between system and measurer and their interactions).
Evidently, the density operators corresponding to these two pure states
will be

\begin{equation}
\begin{cases}
\hat{\rho}_{1} & =\left|a_{1}\right|^{2}\left|+\right\rangle \left\langle +\right|+a_{1}a_{2}^{*}\left|+\right\rangle \left\langle -\right|+a_{1}^{*}a_{2}\left|-\right\rangle \left\langle +\right|+\left|a_{2}\right|^{2}\left|-\right\rangle \left\langle -\right|\\
\hat{\rho}_{2} & =\left|b_{1}\right|^{2}\left|+\right\rangle \left\langle +\right|+b_{1}b_{2}^{*}\left|+\right\rangle \left\langle -\right|+b_{1}^{*}b_{2}\left|-\right\rangle \left\langle +\right|+\left|b_{2}\right|^{2}\left|-\right\rangle \left\langle -\right|
\end{cases}
\end{equation}
or, matricially, in the basis $\left\{ \left|+\right\rangle ,\left|-\right\rangle \right\} $,

\begin{equation}
\hat{\rho}_{1}=\left(\begin{array}{cc}
\left|a_{1}\right|^{2} & a_{1}a_{2}^{*}\\
a_{1}^{*}a_{2} & \left|a_{2}\right|^{2}
\end{array}\right),\:\hat{\rho}_{2}=\left(\begin{array}{cc}
\left|b_{1}\right|^{2} & b_{1}b_{2}^{*}\\
b_{1}^{*}b_{2} & \left|b_{2}\right|^{2}
\end{array}\right).\label{r12}
\end{equation}

In each of the (\ref{r12}), we have the possible results ``$+$''
and ``$-$'' for a measurement perfomed on the system, with the
respective probabilities given by the diagonal elements - called \emph{populations}
- $\left|a_{1}\right|^{2}$ and $\left|a_{2}\right|^{2}$, in case
the system were found in a state $\hat{\rho}_{1}$, or $\left|b_{1}\right|^{2}$
and $\left|b_{2}\right|^{2}$, in case the state were in the state
$\hat{\rho}_{2}$. However, each operator describes a quantum superposition
and the different possibilities/end states of the measurement are
correlated through the non-diagonal elements - the so-called \emph{coherences}:
before the measurement is performed and the wave function suffers
reduction, the systems finds itself in both $\left|+\right\rangle $
\emph{and} $\left|-\right\rangle $ states.

Let us suppose now that in the instant $t_{0}$ we are not sure if
the system is in the state $\left|\psi_{1}\right\rangle $ or $\left|\psi_{2}\right\rangle $,
but we do know that there are probabilities $p_{1}$ and $p_{2}$
of finding the system in either one \emph{or} the other state. The
system will be in that case a statistical mixture described by the
density operator $\hat{\rho}$,

\begin{equation}
\hat{\rho}\left(t_{0}\right)=p_{1}\hat{\rho}_{1}+p_{2}\hat{\rho}_{2}=\left(\begin{array}{ccc}
p_{1}\left|a_{1}\right|^{2}+p_{2}\left|b_{1}\right|^{2} &  & p_{1}a_{1}a_{2}^{*}+p_{2}b_{1}b_{2}^{*}\\
p_{1}a_{1}^{*}a_{2}+p_{2}b_{1}^{*}b_{2} &  & p_{1}\left|a_{2}\right|^{2}+p_{2}\left|b_{2}\right|^{2}
\end{array}\right).
\end{equation}
Once again, the system is in both $\left|+\right\rangle $ \emph{and}
$\left|-\right\rangle $ states prior to the measurement, with probabilities
for the possible results given by the populations and the interferences
between the possible states given by the coherences. There is a difference,
however, if for some reason the coherences disappear in some future
instant $t_{1}$:

\begin{equation}
\hat{\rho}\left(t_{0}\right)=\left(\begin{array}{ccc}
p_{1}\left|a_{1}\right|^{2}+p_{2}\left|b_{1}\right|^{2} &  & p_{1}a_{1}a_{2}^{*}+p_{2}b_{1}b_{2}^{*}\\
p_{1}a_{1}^{*}a_{2}+p_{2}b_{1}^{*}b_{2} &  & p_{1}\left|a_{2}\right|^{2}+p_{2}\left|b_{2}\right|^{2}
\end{array}\right)\longrightarrow\hat{\rho}\left(t_{1}\right)=\left(\begin{array}{ccc}
p_{1}\left|a_{1}\right|^{2}+p_{2}\left|b_{1}\right|^{2} &  & 0\\
0 &  & p_{1}\left|a_{2}\right|^{2}+p_{2}\left|b_{2}\right|^{2}
\end{array}\right).
\end{equation}
In this case, we no longer have interference between the $\left|+\right\rangle $
and $\left|-\right\rangle $ states, they are in a \emph{classical
superposition}: the state has already been defined, but we do not
know which it is, we have only the probabilities given by the populations.
This phenomenon is known as \emph{decoherence}\cite{key-4,key-44,key-46,key-61,key-64,key-65,key-66}.

Quantum mechanics is an inherently statistical theory: it is not possible
to predict the result of a single measurement, we can only make predictions
regarding a series of measurements of an ensemble. Arguably, the state
vector becomes a more fitting description of the post-measurement
system exactly because its ensemble approach does not predict the
future of each system as \emph{completely} as the state vector. \cite{key-37}

As a criterion to determine the instant when the classical world emerged
from the quantum one, we will employ the moment when the quantum interferences
between the possible measurement results disappear, and only classical
indeterminacies remain \cite{key-38}. In other words, we are measuring
the length of time after which the system cannot be represent by a
state vector, the \emph{decoherence time} (in our notation, $t_{1}$).
This is not, however, a universally accepted position \textcolor{blue}{-}
see \cite{key-39}and Conclusion section for some criticism.

\section{Pointer basis}

Now that we have clearly stated the nature of the problem, we can
present its solution, as proposed by Zurek in 1981. \cite{key-6}
In his formulation of the measurement Hamiltonian, von Neumann \cite{key-5}
considers solely a system interacting with a measurement apparatus.
However, a more thorough discussion requires that we take into account
as well the presence of an environment $B$ interacting with the measurement
apparatus $A$ \cite{key-31}. The result of this effect in Everett's
formulation is that we cannot observe macroscopically the measurement
apparatus in a superposition of states/branches. The environment constantly
monitors the apparatus, causing the apparent collapse of the wave
function.

It is worth mentioning that the importance of the environment was
first noted by Zeh during his research on foundations of quantum mechanics.
Indeed, Zeh seems to have reached some of Everett's conclusions independently,
as noted on \cite{key-39}. This made him very enthusiastic about
Everett's formulation and interested in the universal wave function
problem \cite{key-24}, as opposed to orthodox interpretation of quantum
mechanics. On the other hand, Zurek's works on decoherence and environment
effects were more ``practical'' and close to the orthodox interpretation,
although he was developing post-doctoral researches under J. A. Wheeler,
Everett's supervisor. The reference \cite{key-39} shows all the history
and contributions of Zeh and Zurek to the problem.

We will follow here the Zurek's work \cite{key-6}.\textcolor{blue}{{}
}If $\hat{P}_{A}$ is the observable we wish to measure, an ideal
apparatus will leave the system in one of the eigenstates of $\hat{P}_{A}$,
not any relative state, but we have already seen this is not a simple
task. In introducing the environment in the description, Zurek \cite{key-6}
imposed some conditions that had to be satisfied. In his original
article, he admits these conditions are stronger than necessary, and
for this reason here we will keep only two of them:
\begin{enumerate}
\item The environment does not interact with the system (i.e. $\hat{H}_{SB}=0$).
Otherwise, the state of the system would keep suffering environmental
interference after the end of the measurement. (This could mean two
repeated measurements of the same observable could give different
results, which is against the tenets of quantum mechanics.)
\item The system-observer interaction is well-localized in time.
\end{enumerate}
The interaction between the observer and the environment establishes
non-separable correlations between the two, so that this interaction
will only contaminate information obtained from the system. However,
if the interaction commutes with the observable to be measured (apparatus),
this observable will not be perturbed and the measurement will be
trustworthy. For this to happen, the apparatus-environment interaction
must commute with the observable to be measured, that is, $\left[\hat{P}_{A},\hat{H}_{AB}\right]=0$.
The apparatus-environment interaction defines, then, which observable
is to be measured. The basis of eigenstates of $\hat{P}_{A}$ will
contain, then, the possibilities of measurement and will be called
the \emph{pointer basis}. This basis will contain only classical states,
which cannot interfere with each other.

The role of the system $S$ is to determine uniquely the relative
states of the measurer $A$, according to (\ref{medidavN}). Hence,
the two steps of the process of information acquisition described
in Sec. \ref{sub:Basis-ambiguity} will be attributed first to the
system-observer interaction (pre-measurement) and to the observer-environment
interaction (collapse).

There is, however, works on quantum measurement theory where the condition
(1) is not obeyed \cite{key-38,key-40,key-41}.

\subsection{Example}

In Sec. \ref{sec:pre-mes} we analysed the process of \emph{pre-measurement}
and, at this stage, we considered only the main system $S$ and the
measurement apparatus $A$. Let us now approach the complete process,
from the initial evolution of the system until the measurement, step-by-step.
Our analysis involves:
\begin{itemize}
\item the initial evolution of the system $S$ and the measurement apparatus
$A$, without interaction between them;
\item the pre-measurement process, with the interaction between $S$ and
$A$;
\item the beginning of the measurement process, with the introduction of
the environment $B$, interacting with $A$;
\item the determination of the evolution under the effects of the environment;
\item the average of the environmental effects;
\item the end of the measurement process, with analysis of the final $S+A$
state after a long time.
\end{itemize}

\subsubsection{Initial evolution}

Suppose we have the Hamiltonians

\begin{equation}
\hat{H}_{S}=\hbar\omega_{0}\hat{\sigma}_{z}^{\left(S\right)},\:\hat{H}_{A}=\hbar\omega_{0}\hat{\sigma}_{z}^{\left(A\right)}.
\end{equation}

They are both just simple two-level systems, where the energy gap
equals $\omega_{0}$. If the initial state on (\ref{medidavN}) form
evolves only under the influence of these two Hamiltonians, with the
evolution operator

\begin{equation}
\hat{U}_{S+A}\left(t\right)=\exp\left\{ -\frac{i}{\hbar}\left[\hat{H}_{S}+\hat{H}_{A}\right]t\right\} =\exp\left\{ -i\omega_{0}t\left[\hat{\sigma}_{z}^{\left(S\right)}+\hat{\sigma}_{z}^{\left(A\right)}\right]\right\} 
\end{equation}
we will not have the necessary evolution to the pre-measurement process.
To see this in more detail, let us analyse the effect of $\hat{U}_{S+A}\left(t\right)$.
As $\left[\hat{\sigma}_{z}^{\left(S\right)},\hat{\sigma}_{z}^{\left(A\right)}\right]=0$,
the exponential can be split,

\begin{equation}
\hat{U}_{S+A}\left(t\right)=\exp\left[-i\omega_{0}t\hat{\sigma}_{z}^{\left(S\right)}\right]\exp\left[-i\omega_{0}t\hat{\sigma}_{z}^{\left(A\right)}\right]
\end{equation}
which allows us to view the influence of each term over the initial
state

\begin{equation}
\left|\psi\right\rangle _{SA}=\left(s_{+}\left|+\right\rangle +s_{-}\left|-\right\rangle \right)\left|+\right\rangle ,\:\left|s_{+}\right|^{2}+\left|s_{-}\right|^{2}=1.
\end{equation}

The measurement apparatus $A$ just adds a global phase to the initial
state,

\begin{eqnarray}
\hat{U}_{S+A}\left(t\right)\left|\psi\right\rangle _{SA} & = & \exp\left[-i\omega_{0}t\hat{\sigma}_{z}^{\left(S\right)}\right]\exp\left[-i\omega_{0}t\hat{\sigma}_{z}^{\left(A\right)}\right]\left(s_{+}\left|+\right\rangle +s_{-}\left|-\right\rangle \right)\left|+\right\rangle \nonumber \\
 & = & e^{-i\omega_{0}t}\exp\left[-i\omega_{0}t\hat{\sigma}_{z}^{\left(S\right)}\right]\left(s_{+}\left|+\right\rangle +s_{-}\left|-\right\rangle \right)\left|+\right\rangle 
\end{eqnarray}
which has no physical relevance (a global phase vanishes in the density
operator). On the other hand, the system part $S$ on $\hat{U}_{S+A}\left(t\right)$
makes relevant changes (we have omitted the global phase),

\begin{equation}
\hat{U}_{S+A}\left(t\right)\left|\psi\right\rangle _{SA}=\left(s_{+}e^{-i\omega_{0}t}\left|+\right\rangle +s_{-}e^{i\omega_{0}t}\left|-\right\rangle \right)\left|+\right\rangle 
\end{equation}

Then, we have a different system state, but it is not accessible,
for the moment, to the measurement apparatus. Let us suppose that
initially the system evolves under the influence of only the Hamiltonians
$\hat{H}_{S}$ and $\hat{H}_{A}$ for a period $\triangle t$. Then,
if we define $a\equiv s_{+}e^{-i\omega_{0}\triangle t}$ and $b\equiv s_{-}e^{i\omega_{0}\triangle t}$,
we will have our initial state for the pre-measurement process shown
on (\ref{medidavN}):

\begin{equation}
\left|\psi\left(\triangle t\right)\right\rangle _{SA}=\left(a\left|+\right\rangle +b\left|-\right\rangle \right)\left|+\right\rangle .\label{inicial}
\end{equation}

\subsubsection{The pre-measurement process}

Now, we introduce the system - measurement apparatus interaction,
$\hat{H}_{SA}$, as in our example (\ref{HsaZurek}) and perform the
pre-measurement. It is reasonable to suppose that this interaction
is so strong that we can omit the $S$ and $A$ Hamiltonians on the
pre-measurement process, evolving (\ref{inicial}) only under the
evolution operator

\begin{equation}
\hat{U}_{SA}\left(t\right)=\exp\left[-\frac{i}{\hbar}\hat{H}_{SA}t\right]\label{Usa}
\end{equation}
and, if the state (\ref{inicial}) evolves under (\ref{Usa}) on the
\emph{pre-measurement time} (\ref{tpm}), we will have the (\ref{psitm})
state,

\begin{equation}
\hat{U}_{SA}\left(\tau_{pm}\right)\left|\psi\left(\triangle t\right)\right\rangle _{SA}=a\left|++\right\rangle +b\left|--\right\rangle \label{pmstate}
\end{equation}

\subsubsection{Introduction of the environment}

Now, we need to add the environment $B$, which we will dop in the
form of a bath of harmonic oscillators

\begin{eqnarray*}
\hat{H}_{B} & = & \hbar\underset{k}{\sum}\omega_{k}\hat{b}_{k}^{\dagger}\hat{b}_{k},
\end{eqnarray*}
where the $\omega_{k}$ are frequencies, and $\hat{b}_{k}^{\dagger}$
and $\hat{b}_{k}$ are creation and destruction operators for the
bath associated to each mode $k$ The $A-B$ interaction will define
the observable to be measured, as it defines the pointer basis. If
we wish to measure $\hat{P}_{A}=\hat{\sigma}_{z}^{\left(A\right)}$,
we can interact the measurement apparatus with the bath according
to

\begin{equation}
\hat{H}_{AB}=\hbar\hat{\sigma}_{z}^{\left(A\right)}\underset{k}{\sum}\left(g_{k}\hat{b}_{k}^{\dagger}+g_{k}^{*}\hat{b}_{k}\right)\label{HAB}
\end{equation}
where the $g_{k}$ are constant coefficients. This interaction, referred
to in the literature as \emph{phase-damping} \cite{key-36}, causes
decoherence without affecting the populations. 

At this point, we will use the density operator formalism. Let us
consider the environment is in a thermal state that does not suffer
change as time passes:

\begin{equation}
\begin{array}{ccc}
\hat{\rho}_{B} & = & \underset{k}{\prod}\frac{1}{Z_{K}}\exp\left(-\beta\hbar\omega_{k}\hat{b}_{k}^{\dagger}\hat{b}_{k}\right)\\
 & = & \underset{k}{\prod}\frac{1}{Z_{K}}\underset{n}{\sum}\left|n\right\rangle \left\langle n\right|\exp\left(-n\beta\hbar\omega_{k}\right)
\end{array},\:Z_{k}=\frac{1}{1-e^{-\beta\hbar\omega_{k}}}\label{RBTermico}
\end{equation}
($Z_{k}$ is the partition function). Let us to count time from the
pre-measurement process, when the state of $S+A$ (\ref{pmstate})
has the form

\begin{equation}
\hat{\rho}_{SA}\left(0\right)=\left(\begin{array}{cccc}
\left|a\right|^{2} & 0 & 0 & ab^{*}\\
0 & 0 & 0 & 0\\
0 & 0 & 0 & 0\\
a^{*}b & 0 & 0 & \left|b\right|^{2}
\end{array}\right)
\end{equation}

When the measurement begins - immediately after the pre-measurement
- the environment had not interacted with the system and the measurement
apparatus yet. For this reason, we will consider the initial state
on the form
\begin{equation}
\hat{\rho}_{SAB}\left(0\right)=\hat{\rho}_{SA}\left(0\right)\otimes\hat{\rho}_{B}\left(0\right)\label{rsabzero}
\end{equation}
with $\hat{\rho}_{B}\left(0\right)$ being the thermal state. 

As mentioned above, (\ref{HAB}) causes decoherence, so that we expect
$\hat{\rho}_{SA}\left(0\right)$ to evolve to a state of the form

\begin{equation}
\hat{\rho}_{SA}\left(t\right)=\left(\begin{array}{cccc}
\left|a\right|^{2} & 0 & 0 & 0\\
0 & 0 & 0 & 0\\
0 & 0 & 0 & 0\\
0 & 0 & 0 & \left|b\right|^{2}
\end{array}\right)
\end{equation}
where we have a statistical mixture, that is, an indefinition in the
state of the system, that has probability $\left|a\right|^{2}$ of
being at $\left|++\right\rangle $ and $\left|b\right|^{2}$ of being
at $\left|--\right\rangle $. To verify this, we must calculate the
density operator referring to the Hamiltonian

\begin{equation}
\hat{H}_{total}=\hbar\omega_{0}\left[\hat{\sigma}_{z}^{\left(S\right)}+\hat{\sigma}_{z}^{\left(A\right)}\right]+\hbar\hat{\sigma}_{z}^{\left(A\right)}\underset{k}{\sum}\left(g_{k}\hat{b}_{k}^{\dagger}+g_{k}^{*}\hat{b}_{k}\right)+\hbar\underset{k}{\sum}\omega_{k}\hat{b}_{k}^{\dagger}\hat{b}_{k}\label{Hsab}
\end{equation}

It is important to note on (\ref{Hsab}), we are obeying the two previous
Zurek's hypothesis, i.e., we have not considered an interaction between
the system$S$ and the bath $B$ (hypothesis 1) and, as the $\hat{H}_{SA}$
is strong and well-localized in time, it is not considered (hypothesis
2). As we are not considering any other conditions to the $\hat{H}_{AB}$
for its time duration and particularly its intensity compared with
$\hat{H}_{S}$ and $\hat{H}_{A}$, then these Hamiltonians are considered
on the subsequent evolution too, as well $\hat{H}_{B}$.

\subsubsection{Evolution due to the environment}

Following a procedure analogous to the one seen in \cite{key-42},
we find (see Appendix A) the evolution operator for (\ref{Hsab})
is:

\begin{equation}
\hat{U}_{total}\left(t\right)=\Phi\left(t\right)e^{-i\omega_{0}t\left[\hat{\sigma}_{z}^{\left(S\right)}+\hat{\sigma}_{z}^{\left(A\right)}\right]}e^{\hat{\sigma}_{z}^{\left(A\right)}\underset{k}{\sum}\left[g_{k}\hat{b}_{k}^{\dagger}\varphi_{k}\left(t\right)-g_{k}^{*}\hat{b}_{k}\varphi_{k}^{*}\left(t\right)\right]}\label{U1}
\end{equation}
where

\begin{equation}
\Phi\left(t\right)\equiv e^{-\underset{k}{\sum}\frac{\left|g_{k}\right|^{2}}{\omega_{k}}\varphi_{k}^{*}\left(t\right)}e^{it\underset{k}{\sum}\frac{\left|g_{k}\right|^{2}}{\omega_{k}}}e^{\frac{1}{2}\underset{k}{\sum}\left|g_{k}\right|^{2}\left|\varphi_{k}\left(t\right)\right|^{2}}
\end{equation}

The density operator at a posterior time will be given by:

\begin{equation}
\hat{\rho}_{SAB}\left(t\right)=\hat{U}_{SAB}\left(t\right)\hat{\rho}_{SAB}\left(0\right)\hat{U}_{SAB}^{\dagger}\left(t\right)
\end{equation}
According to (\ref{U1}), $\hat{U}_{total}^{\dagger}\left(t\right)$
is given by

\begin{equation}
\hat{U}_{total}^{\dagger}\left(t\right)=\Phi^{*}\left(t\right)e^{i\omega_{0}t\left[\hat{\sigma}_{z}^{\left(S\right)}+\hat{\sigma}_{z}^{\left(A\right)}\right]}e^{-\hat{\sigma}_{z}^{\left(A\right)}\underset{k}{\sum}\left[g_{k}\hat{b}_{k}^{\dagger}\varphi_{k}\left(t\right)-g_{k}^{*}\hat{b}_{k}\varphi_{k}^{*}\left(t\right)\right]}.
\end{equation}
Then, considering the initial state form (\ref{rsabzero}),

\begin{eqnarray}
\hat{\rho}_{SAB}\left(t\right) & = & \left|\Phi\left(t\right)\right|^{2}e^{-i\omega_{0}t\left[\hat{\sigma}_{z}^{\left(S\right)}+\hat{\sigma}_{z}^{\left(A\right)}\right]}e^{\hat{\sigma}_{z}^{\left(A\right)}\underset{k}{\sum}\left[g_{k}\hat{b}_{k}^{\dagger}\varphi_{k}\left(t\right)-g_{k}^{*}\hat{b}_{k}\varphi_{k}^{*}\left(t\right)\right]}\hat{\rho}_{SA}\left(0\right)\hat{\rho}_{B}\left(0\right)\nonumber \\
 & \times & e^{i\omega_{0}t\left[\hat{\sigma}_{z}^{\left(S\right)}+\hat{\sigma}_{z}^{\left(A\right)}\right]}e^{-\hat{\sigma}_{z}^{\left(A\right)}\underset{k}{\sum}\left[g_{k}\hat{b}_{k}^{\dagger}\varphi_{k}\left(t\right)-g_{k}^{*}\hat{b}_{k}\varphi_{k}^{*}\left(t\right)\right]}
\end{eqnarray}

\subsubsection{Averaging over the environmental degrees of freedom}

If we are interested in the effect of the environment during a time
interval when the measurement occurs, then we must consider a time
average of that effect. Statistically, on the density operator formalism,
this average is made taking the partial trace over the environmental
degrees of freedom. Doing this operation, we find (see Appendix B):

\begin{eqnarray*}
\left\langle s_{p}a_{p}\right|\hat{\rho}_{S}\left(t\right)\left|s_{q}a_{q}\right\rangle  & = & \left|\Phi\left(t\right)\right|^{2}e^{-i\omega_{0}t\left(s_{p}+a_{p}\right)}e^{i\omega_{0}t\left(s_{q}+a_{q}\right)}\\
 & \times & \left\langle s_{p}a_{p}\right|\hat{\rho}_{S}\left(0\right)\left|s_{q}a_{q}\right\rangle e^{-\frac{\left(a_{p}-a_{q}\right)}{2}^{2}\underset{k}{\sum}\left|g_{k}\right|^{2}\left|\varphi_{k}\left(t\right)\right|^{2}\coth\left(\frac{\beta\hbar\omega_{k}}{2}\right)}
\end{eqnarray*}

It remains to take the continuous limit by defining a spectral density
\cite{key-36}

\begin{equation}
J\left(\omega\right)=\underset{k}{\sum}\left|g_{k}\right|^{2}\delta\left(\omega-\omega_{k}\right)
\end{equation}

The final result will be

\begin{eqnarray}
\left\langle s_{p}a_{p}\right|\hat{\rho}_{SA}\left(t\right)\left|s_{q}a_{q}\right\rangle  & = & e^{-i\omega_{0}t\left(s_{p}-s_{q}\right)}e^{-i\omega_{0}t\left(a_{p}-a_{q}\right)}\nonumber \\
 & \times & e^{-\left(a_{p}-a_{q}\right)^{2}\int_{0}^{\infty}d\omega J\left(\omega\right)\frac{1-\cos\left(\omega t\right)}{\omega^{2}}\coth\left(\frac{\beta\hbar\omega}{2}\right)}\left\langle s_{p}a_{p}\right|\hat{\rho}_{SA}\left(0\right)\left|s_{q}a_{q}\right\rangle \label{rs3}
\end{eqnarray}

Finding the matrix elements involves, then, the integral:

\begin{equation}
I_{1}\left(t\right)=\int_{0}^{\infty}d\omega J\left(\omega\right)\frac{1-\cos\left(\omega t\right)}{\omega^{2}}\coth\left(\frac{\beta\hbar\omega}{2}\right)\label{integrais}
\end{equation}
Re-writing (\ref{rs3}) using that definition

\begin{equation}
\left\langle s_{p}a_{p}\right|\hat{\rho}_{SA}\left(t\right)\left|s_{q}a_{q}\right\rangle =e^{-i\omega_{0}t\left(s_{p}-s_{q}\right)}e^{-i\omega_{0}t\left(a_{p}-a_{q}\right)}e^{-\left(a_{p}-a_{q}\right)^{2}I_{1}\left(t\right)}\left\langle s_{p}a_{p}\right|\hat{\rho}_{SA}\left(0\right)\left|s_{q}a_{q}\right\rangle \label{rs4}
\end{equation}
that is,

\begin{equation}
\hat{\rho}_{SA}\left(t\right)=\left(\begin{array}{cccc}
\left|a\right|^{2} & 0 & 0 & e^{-i4\omega_{0}t}e^{-4I_{1}\left(t\right)}ab^{*}\\
0 & 0 & 0 & 0\\
0 & 0 & 0 & 0\\
e^{i4\omega_{0}t}e^{-4I_{1}\left(t\right)}a^{*}b & 0 & 0 & \left|b\right|^{2}
\end{array}\right)\label{roSt}
\end{equation}

\subsubsection{End of the measurement process}

It must be said that, in the limit when $t\rightarrow0$, the initial
condition is re-obtained. As time passes, the exponential of $I_{1}\left(t\right)$
in (\ref{roSt}) makes the off-diagonal non-zero elements vanish:
\begin{equation}
\hat{\rho}_{SA}\left(t\right)\longrightarrow\left(\begin{array}{cccc}
\left|a\right|^{2} & 0 & 0 & 0\\
0 & 0 & 0 & 0\\
0 & 0 & 0 & 0\\
0 & 0 & 0 & \left|b\right|^{2}
\end{array}\right).
\end{equation}

In accordance to our discussion in Sec. \ref{sub:probabilities},
we can say that we experience the system as if it were a classical
mixture.

It remains to be evaluated the problem discussed previously, the basis
ambiguity. Let us rewrite the final state in the basis $\left\{ \left|++\right\rangle _{xy},\left|+-\right\rangle _{xy},\left|-+\right\rangle _{xy},\left|--\right\rangle _{xy}\right\} $:

\begin{equation}
\hat{\rho}_{SA}\left(t\right)=\left(\begin{array}{ccccccc}
1 &  & 2\left|a\right|^{2}-1 &  & 2\left|a\right|^{2}-1 &  & 1\\
2\left|a\right|^{2}-1 &  & 1 &  & 1 &  & 2\left|a\right|^{2}-1\\
2\left|a\right|^{2}-1 &  & 1 &  & 1 &  & 2\left|a\right|^{2}-1\\
1 &  & 2\left|a\right|^{2}-1 &  & 2\left|a\right|^{2}-1 &  & 1
\end{array}\right)
\end{equation}
(where we employed the identity $\left|a\right|^{2}+\left|b\right|^{2}=1$).

We see in this case that there is no possible choice of the coefficients
$x$ or $y$ that defines another basis in which this state could
also represent a pre-measurement - the pointer basis is the only one
where this occurs. Hence, the problem of basis ambiguity is solved.

\section{Decoherence and beyond}

In the following year of the pointer states paper,\textcolor{red}{{}
}another article by Zurek comes to light, suggesting that the process
of monitoring by the environment was responsible for the apparent
wave-function collapse. \cite{key-43}

The disappearance of certain the interference terms that give rise
to a preferred eigenbasis of a quantum observable to become stable,
while the others vanish, was dubbed \emph{decoherence. }The term was
popularized in a Physics Today article \cite{key-44}, and became
a major focus of research in the decades to come \textcolor{red}{\cite{key-4,key-39,key-45,key-46,key-47,key-61,key-62,key-63,key-64,key-65,key-66}}.
With the emergence of studies in quantum information theory, the decoherence
became a real problem to be dealt with, as the interference terms
are instrumental in the processing of quantum bits. \cite{key-4,qc1}
Therefore, any process that eliminates these interferences can be
seen as detrimental, prompting the emergence of a series of schemes
to tackle decoherence such as quantum codes \cite{qc1} and decoherence-free
subspaces \cite{qc2}.

The \emph{einselection \cite{key-43} }brought about by the constant
monitoring of the environment, that is, the choice of a preferred
basis by these means, was seen as the solution to many conundrums,
including Schrödinger's cat -- if the box is not isolated, environment-induced
decoherence would be responsible for causing the cat to show up in
one of the two classical states \cite{key-45}.

The further discovery that the pointer states were not only the ones
selected by the kind of interaction with the environment, but also
the ones whose information spreads more easily across the neighboring
system (an analogy between competing bases and species led to the
phrase ``quantum Darwinism'' \cite{key-77,key-72,key-73,key-74,key-75,key-76}),
allowing a large number of observers to reach the same conclusion
about which pointer basis was selected, hinted at a mechanism that
allowed the emergence of classical macroscopic reality from the the
quantum microscopic world. \cite{key-4} This was the rise of a ``new
orthodoxy'' in quantum mechanics or, as Zurek calls it, an ``existential
interpretation'' \cite{key-77,key-32} which he fashions as somewhere
midway between the original Bohr and Everett interpretations \cite{key-46}.
If for a long while most physics acted as if there was no problem
to be solved, now many were confident that it had been settled for
good. For example, Tegmark and Wheeler \cite{key-62} suggested that
\textquotedbl{}it is time to update the quantum textbooks\textquotedbl{}
against the wave function collapse.

\textcolor{black}{However, this optimism is not unanimous within the
academic community. Other approaches to the problem of measurement
in quantum mechanics that have great acceptance are: the quantum state
diffusion \cite{citation-1}; dynamical reduction models \cite{citation-2}
like the Ghirardi-Rimini-Weber original proposal (GRW) \cite{citation-3}
and related developments \cite{citation-4,citation-5}; and the superselection
rules \cite{key-79,key-80,key-81,key-82,citation-6}. Thus, quantum
measurement theory is still a field open for debate.}

A common criticism of the decoherence program is that it does not
explain collapse, only the transformation of a quantum pure superposition
state into a quantum mixture. Therefore, a partial density matrix
would never be capable of describing the collapse of a single quantum
system described by a state vector, because collapse would be implied
in appealing to ensemble description \cite{key-47}. Others note that
the decoherence project would be circular if the loss of coherence
depended on the appearance of a sum of random phases in the off-diagonal
terms, because the very existence of these would depend on a non-unitary
evolution of the universe. \cite{key-48}

Zurek, nevertheless, has made an effort to address such criticism,
for example by deriving Born's rule within the decoherence framework
\cite{BORNrule01}, as we explain in Appendix D. Even so, this derivation
has arguably the shortcoming of being incompatible with the usual
version of Everett's interpretation \cite{key-27}, and would require
further require further assumptions to properly explain the origin
of probabilities in a branching universe. In the above development
of this article, we have assumed an Everettian framework in which
there is no collapse of the wave function, just a subjective impression
that it has happened, and therefore this is all that the decoherence
explain in this scenario. But we have not tackled the problem of the
meaning of Probabilities mentioned in Sec. 2.1 When delving into that,
it might become necessary to introduce new assumptions into the theory:
maybe postulates to explain Born's law, or even a collapse postulate.

For both supporters and detractors, however, the decoherence edifice
built from the pointer basis problem tackled in 1981 has assumed such
great proportions that it cannot be ignored.

\section{Conclusion}

In Sec. 2.1 we listed two classes of problems related to Everett's
interpretation. The pointer-basis project in 1981 was presented to
solve simply one specific point of that second class of problems:
the lack of quantum interference effects on macro scale. Sec. 5 furnishes
a short discussion about some implications after the pointer-basis
on the context of decoherence. However, it is important to notice
that the concept of pointer states does not necessarily imply an acceptance
of decoherence as the solution of all problems of quantum measurement,
or of any of the other theories mentioned in the previous section.
As we said previously, pointer states are simply eigenstates of the
observable of the measurement apparatus that represent the possible
positions of the display pointer of the equipment. Despite being motivated
by Everett's relative state interpretation, pointer states presume
simply an interaction between the measurement apparatus and the environment
and that the observable to be measured commutes with this interaction.
The study of pointer states allows not only a better understanding
of a broad (and still controversial) field of interpretations of quantum
mechanics, but also to follow concrete implementations in quantum
theory of information - a recent example of which can be found at
\cite{key-78}. We hope this article can help students interested
in both approaches, those who seek a more pragmatic angle, and those
who want to comprehend the fundamental implications of decoherence.

\section*{Acknowledgements}

The authors wish to thank R. d. J. Napolitano for reading and criticizing
the preliminary versions of this paper. Additionally, C. A. Brasil
acknowledges support from Fundação de Amparo à Pesquisa do Estado
de São Paulo (FAPESP), project number 2011/19848-4, Brazil, and A.
O. Caldeira for his hospitality and useful discussions; L. A. de Castro
acknowledges support from the Coordenação de Aperfeiçoamento de Pessoal
de Nível Superior (CAPES).

\section*{Appendix A: Finding the Time Evolution}

We want to discover the unitary evolution operator corresponding to
the following Hamiltonian:

\begin{equation}
\hat{H}_{total}=\hbar\omega_{0}\left[\hat{\sigma}_{z}^{\left(S\right)}+\hat{\sigma}_{z}^{\left(A\right)}\right]+\hbar\hat{\sigma}_{z}^{\left(A\right)}\underset{k}{\sum}\left(g_{k}\hat{b}_{k}^{\dagger}+g_{k}^{*}\hat{b}_{k}\right)+\hbar\underset{k}{\sum}\omega_{k}\hat{b}_{k}^{\dagger}\hat{b}_{k}
\end{equation}

The calculations are based on\cite{key-42}. First, we perform a transformation
to eliminate the last term

\begin{equation}
\hat{U}_{0}=\exp\left(-it\underset{k}{\sum}\omega_{k}\hat{b}_{k}^{\dagger}\hat{b}_{k}\right)
\end{equation}
that is,

\begin{eqnarray*}
\hat{H}_{1} & = & \hat{U}_{0}^{\dagger}\left\{ \hbar\omega_{0}\left[\hat{\sigma}_{z}^{\left(S\right)}+\hat{\sigma}_{z}^{\left(A\right)}\right]+\hbar\hat{\sigma}_{z}^{\left(A\right)}\underset{k}{\sum}\left(g_{k}\hat{b}_{k}^{\dagger}+g_{k}^{*}\hat{b}_{k}\right)\right\} \hat{U}_{0}
\end{eqnarray*}

As $\left[\hat{U}_{0},\hat{\sigma}_{z}^{\left(S\right)}\right]=\left[\hat{U}_{0},\hat{\sigma}_{z}^{\left(A\right)}\right]=0$,
the term $\hbar\omega_{0}\left[\hat{\sigma}_{z}^{\left(S\right)}+\hat{\sigma}_{z}^{\left(A\right)}\right]$
will not be altered. Therefore, 

\begin{eqnarray*}
\hat{H}_{1} & = & \hbar\omega_{0}\left[\hat{\sigma}_{z}^{\left(S\right)}+\hat{\sigma}_{z}^{\left(A\right)}\right]+\hbar\hat{\sigma}_{z}^{\left(A\right)}\hat{U}_{0}^{\dagger}\underset{k}{\sum}\left(g_{k}\hat{b}_{k}^{\dagger}+g_{k}^{*}\hat{b}_{k}\right)\hat{U}_{0}
\end{eqnarray*}
Let us focus on the term $\hat{U}_{0}^{\dagger}\underset{k}{\sum}g_{k}\hat{b}_{k}^{\dagger}\hat{U}_{0}$,
given that its conjugate will yield the remaining term. In the first
place, we will have the commutation relations:

\begin{eqnarray*}
\left[\hat{b}_{k},\hat{b}_{k'}^{\dagger}\right] & = & \delta_{k,k'}\\
\left[\hat{b}_{k},\hat{b}_{k'}\right]=\left[\hat{b}_{k}^{\dagger},\hat{b}_{k'}^{\dagger}\right] & = & 0
\end{eqnarray*}
hence, all the terms in $\hat{U}_{0}$ referring to $k'\neq k$ will
cancel out with its equivalents in $\hat{U}_{0}^{\dagger}$, leaving
only those with$k'=k$:

\begin{equation}
\hat{U}_{0}^{\dagger}\underset{k}{\sum}g_{k}\hat{b}_{k}^{\dagger}\hat{U}_{0}=\underset{k}{\sum}exp\left(it\omega_{k}\hat{b}_{k}^{\dagger}\hat{b}_{k}\right)g_{k}\hat{b}_{k}^{\dagger}exp\left(-it\omega_{k}\hat{b}_{k}^{\dagger}\hat{b}_{k}\right)\label{apaux1}
\end{equation}
To calculate (\ref{apaux1}), we will employ the following relation
involving the operators $\hat{A}$ and $\hat{B}$ \cite{key-49,key-50},
which can be obtained simply by expanding its exponentials in power
series and grouping together similar terms:

\begin{equation}
e^{\hat{A}}\hat{B}e^{-\hat{A}}=\hat{B}+\left[\hat{A},\hat{B}\right]+\frac{1}{2!}\left[\hat{A},\left[\hat{A},\hat{B}\right]\right]+...\label{WILCOX1}
\end{equation}
Choosing $\hat{A}=it\omega_{k}\hat{b}_{k}^{\dagger}\hat{b}_{k}$ and
$\hat{B}=g_{k}\hat{b}_{k}^{\dagger}$, we find

\begin{equation}
\hat{U}_{0}^{\dagger}\underset{k}{\sum}g_{k}\hat{b}_{k}^{\dagger}\hat{U}_{0}=\underset{k}{\sum}g_{k}\hat{b}_{k}^{\dagger}e^{i\omega_{k}t}
\end{equation}
and, finally,

\begin{eqnarray}
\hat{H}_{1} & = & \hbar\omega_{0}\left[\hat{\sigma}_{z}^{\left(S\right)}+\hat{\sigma}_{z}^{\left(A\right)}\right]+\hbar\hat{\sigma}_{z}^{\left(A\right)}\underset{k}{\sum}\left(g_{k}\hat{b}_{k}^{\dagger}e^{i\omega_{k}t}+g_{k}^{*}\hat{b}_{k}e^{-i\omega_{k}t}\right)\label{H1}
\end{eqnarray}

We need, now, to obtain the evolution operator referring to (\ref{H1}).
The difficulty here is that $\hat{H}_{1}$ does not commute with itself
for different times (in other words, $\left[\hat{H}_{1}\left(t\right),\hat{H}_{1}\left(t'\right)\right]\neq0$
when $t\neq t$') because of the time-dependent term $\hbar\hat{\sigma}_{z}^{\left(A\right)}\underset{k}{\sum}\left(g_{k}\hat{b}_{k}^{\dagger}e^{i\omega_{k}t}+g_{k}^{*}\hat{b}_{k}e^{-i\omega_{k}t}\right)$.
For this reason the exponential of $\hat{H}_{1}$ will require the
time-ordering operator $\hat{T}$ \cite{key-51,key-52}. Hence,

\begin{eqnarray}
\hat{U}_{total}\left(t\right) & = & \hat{T}\exp\left[-\frac{i}{\hbar}\int_{0}^{t}\hat{H}_{1}\left(t'\right)dt'\right]\nonumber \\
 & = & \exp\left\{ -i\omega_{0}t\left[\hat{\sigma}_{z}^{\left(S\right)}+\hat{\sigma}_{z}^{\left(A\right)}\right]\right\} \nonumber \\
 & \times & \hat{T}\exp\left[-i\hat{\sigma}_{z}^{\left(A\right)}\underset{k}{\sum}\int_{0}^{t}dt'\left(g_{k}\hat{b}_{k}^{\dagger}e^{i\omega_{k}t'}+g_{k}^{*}\hat{b}_{k}e^{-i\omega_{k}t'}\right)\right].\label{U1inicial}
\end{eqnarray}
In this case, it can be useful to employ the result \cite{key-53},

\begin{eqnarray}
 & \hat{T}\exp\left[-i\underset{k}{\sum}\lambda\int_{0}^{t}dt'\left(\hat{a}_{k}^{\dagger}e^{i\omega_{k}t'}+\hat{a}_{k}e^{-i\omega_{k}t'}\right)\right]\nonumber \\
 & =e^{\underset{k}{\sum}\hat{a}_{k}^{\dagger}\phi_{k'}\left(t\right)}\hat{T}\exp\left[-i\underset{k}{\sum}\lambda\int_{0}^{t}dt'e^{-\underset{k'}{\sum}\hat{a}_{k'}^{\dagger}\phi_{k'}\left(t'\right)}\hat{a}_{k}e^{\underset{k'}{\sum}\hat{a}_{k'}^{\dagger}\phi_{k'}\left(t'\right)}e^{-i\omega_{k}t'}\right]\label{Mahan1}
\end{eqnarray}
where

\begin{equation}
\phi_{k}\left(t\right)=-i\lambda\int_{0}^{t}dt'e^{i\omega_{k}t}=\frac{\lambda}{\omega_{k}}\left(1-e^{i\omega_{k}t}\right)
\end{equation}
or

\begin{equation}
\phi_{k}\left(t\right)=\lambda\varphi_{k}\left(t\right),\:\varphi_{k}\left(t\right)=\frac{1-e^{i\omega_{k}t}}{\omega_{k}}
\end{equation}

To apply (\ref{Mahan1}) in (\ref{U1inicial}), we must recognize
the parallels:

\begin{equation}
\begin{cases}
\lambda & \rightarrow\hat{\sigma}_{z}^{\left(2\right)}\\
\hat{a}_{k} & \rightarrow g_{k}^{*}\hat{b}_{k}
\end{cases}
\end{equation}
to obtain

\begin{eqnarray}
\hat{U}_{total}\left(t\right) & = & \exp\left\{ -i\omega_{0}t\left[\hat{\sigma}_{z}^{\left(S\right)}+\hat{\sigma}_{z}^{\left(A\right)}\right]\right\} e^{\hat{\sigma}_{z}^{\left(A\right)}\underset{k}{\sum}g_{k}\hat{b}_{k}^{\dagger}\varphi_{k}\left(t'\right)}\nonumber \\
 & \times & \hat{T}\exp\left[-i\hat{\sigma}_{z}^{\left(A\right)}\underset{k}{\sum}\int_{0}^{t}dt'e^{-\hat{\sigma}_{z}^{\left(A\right)}\underset{k'}{\sum}g_{k'}\hat{b}_{k'}^{\dagger}\varphi_{k'}\left(t'\right)}g_{k}^{*}\hat{b}_{k}e^{\hat{\sigma}_{z}^{\left(A\right)}\underset{k'}{\sum}g_{k'}\hat{b}_{k'}^{\dagger}\varphi_{k'}\left(t'\right)}e^{-i\omega_{k}t'}\right]\nonumber \\
\end{eqnarray}
but, once again, employing (\ref{WILCOX1}), we can re-write the integrand
as

\begin{equation}
e^{-\hat{\sigma}_{z}^{\left(A\right)}\underset{k'}{\sum}g_{k'}\hat{b}_{k'}^{\dagger}\varphi_{k'}\left(t\right)}g_{k}^{*}\hat{b}_{k}e^{\hat{\sigma}_{z}^{\left(A\right)}\underset{k'}{\sum}g_{k'}\hat{b}_{k'}^{\dagger}\varphi_{k'}\left(t\right)}=g_{k}^{*}\left[\hat{b}_{k}+\hat{\sigma}_{z}^{\left(A\right)}g_{k}\varphi_{k}\left(t\right)\right]
\end{equation}
obtaining, after a few manipulations,

\begin{eqnarray}
\hat{U}_{total}\left(t\right) & = & \exp\left(-\underset{k}{\sum}\frac{\left|g_{k}\right|^{2}}{\omega_{k}}\varphi_{k}^{*}\left(t\right)\right)\exp\left(it\underset{k}{\sum}\frac{\left|g_{k}\right|^{2}}{\omega_{k}}\right)\nonumber \\
 & \times & \exp\left\{ -i\omega_{0}t\left[\hat{\sigma}_{z}^{\left(S\right)}+\hat{\sigma}_{z}^{\left(A\right)}\right]\right\} e^{\hat{\sigma}_{z}^{\left(A\right)}\underset{k}{\sum}g_{k}\hat{b}_{k}^{\dagger}\varphi_{k}\left(t\right)}e^{-\hat{\sigma}_{z}^{\left(A\right)}\underset{k}{\sum}g_{k}^{*}\hat{b}_{k}\varphi_{k}^{*}\left(t\right)}
\end{eqnarray}

It is convenient to unite the two exponentials that have operators
of the bath. We will employ another result of \cite{key-49}, valid
for two operators $\hat{A}$ and $\hat{B}$ such that $\left[\hat{A},\left[\hat{A},\hat{B}\right]\right]=\left[\hat{B},\left[\hat{A},\hat{B}\right]\right]=0$,

\begin{equation}
e^{\hat{A}}e^{\hat{B}}=e^{\hat{A}+\hat{B}}e^{\frac{\left[\hat{A},\hat{B}\right]}{2}}\label{BCH}
\end{equation}
This is the Baker-Campbell-Hausdorff formula \cite{key-54,key-55,key-56}.

Choosing $\begin{cases}
\hat{A}= & \hat{\sigma}_{z}^{\left(A\right)}\underset{k}{\sum}g_{k}\hat{b}_{k}^{\dagger}\varphi_{k}\left(t\right)\\
\hat{B}= & -\hat{\sigma}_{z}^{\left(A\right)}\underset{k}{\sum}g_{k}^{*}\hat{b}_{k}\varphi_{k}^{*}\left(t\right)
\end{cases}$, as $\left[\hat{A},\hat{B}\right]=\underset{k}{\sum}\left|g_{k}\right|^{2}\left|\varphi_{k}\left(t\right)\right|^{2}$,
we obtain
\begin{equation}
\hat{U}_{total}\left(t\right)=e^{-\underset{k}{\sum}\frac{\left|g_{k}\right|^{2}}{\omega_{k}}\varphi_{k}^{*}\left(t\right)}e^{it\underset{k}{\sum}\frac{\left|g_{k}\right|^{2}}{\omega_{k}}}e^{\frac{1}{2}\underset{k}{\sum}\left|g_{k}\right|^{2}\left|\varphi_{k}\left(t\right)\right|^{2}}e^{-i\omega_{0}t\left[\hat{\sigma}_{z}^{\left(S\right)}+\hat{\sigma}_{z}^{\left(A\right)}\right]}e^{\hat{\sigma}_{z}^{\left(A\right)}\underset{k}{\sum}\left[g_{k}\hat{b}_{k}^{\dagger}\varphi_{k}\left(t\right)-g_{k}^{*}\hat{b}_{k}\varphi_{k}^{*}\left(t\right)\right]}
\end{equation}

To simplify, we define

\begin{equation}
\Phi\left(t\right)\equiv e^{-\underset{k}{\sum}\frac{\left|g_{k}\right|^{2}}{\omega_{k}}\varphi_{k}^{*}\left(t\right)}e^{it\underset{k}{\sum}\frac{\left|g_{k}\right|^{2}}{\omega_{k}}}e^{\frac{1}{2}\underset{k}{\sum}\left|g_{k}\right|^{2}\left|\varphi_{k}\left(t\right)\right|^{2}}
\end{equation}
so in the end we find
\begin{equation}
\hat{U}_{total}\left(t\right)=\Phi\left(t\right)e^{-i\omega_{0}t\left[\hat{\sigma}_{z}^{\left(S\right)}+\hat{\sigma}_{z}^{\left(A\right)}\right]}e^{\hat{\sigma}_{z}^{\left(A\right)}\underset{k}{\sum}\left[g_{k}\hat{b}_{k}^{\dagger}\varphi_{k}\left(t\right)-g_{k}^{*}\hat{b}_{k}\varphi_{k}^{*}\left(t\right)\right]}.\label{U1-1}
\end{equation}

\section*{Appendix B: A Inner Product}

Once we have an expression for the operator without an explicit time
ordering

\begin{equation}
\hat{U}_{1}\left(t\right)=\Phi\left(t\right)e^{-i\omega_{0}t\left[\hat{\sigma}_{z}^{\left(S\right)}+\hat{\sigma}_{z}^{\left(A\right)}\right]}e^{\hat{\sigma}_{z}^{\left(A\right)}\underset{k}{\sum}\left[g_{k}\hat{b}_{k}^{\dagger}\varphi_{k}\left(t\right)-g_{k}^{*}\hat{b}_{k}\varphi_{k}^{*}\left(t\right)\right]}\label{U1-2}
\end{equation}
where
\begin{equation}
\Phi\left(t\right)\equiv e^{-\underset{k}{\sum}\frac{\left|g_{k}\right|^{2}}{\omega_{k}}\varphi_{k}^{*}\left(t\right)}e^{it\underset{k}{\sum}\frac{\left|g_{k}\right|^{2}}{\omega_{k}}}e^{\frac{1}{2}\underset{k}{\sum}\left|g_{k}\right|^{2}\left|\varphi_{k}\left(t\right)\right|^{2}}
\end{equation}
(see Appendix A), we take the partial trace over the degrees of freedom
of the environment. This will be done considering that the initial
density operator is in a factorizable form

\begin{equation}
\hat{\rho}_{SAB}\left(0\right)=\hat{\rho}_{SA}\left(0\right)\otimes\hat{\rho}_{B}\left(0\right)
\end{equation}
with $\hat{\rho}_{B}\left(0\right)$ being thermal state (\ref{RBTermico}).

The density operator in any posterior time will be given by

\begin{equation}
\hat{\rho}_{SAB}\left(t\right)=\hat{U}_{1}\left(t\right)\hat{\rho}_{SAB}\left(0\right)\hat{U}_{1}^{\dagger}\left(t\right).
\end{equation}
According to (\ref{U1-2}), $\hat{U}_{1}^{\dagger}\left(t\right)$
is given by

\begin{equation}
\hat{U}_{1}^{\dagger}\left(t\right)=\Phi^{*}\left(t\right)e^{i\omega_{0}t\left[\hat{\sigma}_{z}^{\left(S\right)}+\hat{\sigma}_{z}^{\left(A\right)}\right]}e^{-\hat{\sigma}_{z}^{\left(A\right)}\underset{k}{\sum}\left[g_{k}\hat{b}_{k}^{\dagger}\varphi_{k}\left(t\right)-g_{k}^{*}\hat{b}_{k}\varphi_{k}^{*}\left(t\right)\right]}
\end{equation}
hence,

\begin{eqnarray}
\hat{\rho}_{SB}\left(t\right) & = & \left|\Phi\left(t\right)\right|^{2}e^{-i\omega_{0}t\left[\hat{\sigma}_{z}^{\left(S\right)}+\hat{\sigma}_{z}^{\left(A\right)}\right]}e^{\hat{\sigma}_{z}^{\left(A\right)}\underset{k}{\sum}\left[g_{k}\hat{b}_{k}^{\dagger}\varphi_{k}\left(t\right)-g_{k}^{*}\hat{b}_{k}\varphi_{k}^{*}\left(t\right)\right]}\hat{\rho}_{SA}\left(0\right)\hat{\rho}_{B}\left(0\right)\nonumber \\
 & \times & e^{i\omega_{0}t\left[\hat{\sigma}_{z}^{\left(S\right)}+\hat{\sigma}_{z}^{\left(A\right)}\right]}e^{-\hat{\sigma}_{z}^{\left(A\right)}\underset{k}{\sum}\left[g_{k}\hat{b}_{k}^{\dagger}\varphi_{k}\left(t\right)-g_{k}^{*}\hat{b}_{k}\varphi_{k}^{*}\left(t\right)\right]}
\end{eqnarray}

For simplicity, let us consider first the matrix elements referring
to the main system and the measurement apparatus. Following the notation
from Sec. \ref{sec:A-S} , let us consider the basis $\left\{ \left|s_{p}\right\rangle _{S}\left|a_{q}\right\rangle _{A}\right\} \equiv\left\{ \left|s_{p}a_{q}\right\rangle \right\} $
(in a simplified notation) in the joint space $S+A$:

\begin{eqnarray*}
\left\langle s_{p}a_{p}\right|\hat{\rho}_{SB}\left(t\right)\left|s_{q}a_{q}\right\rangle  & = & \left\langle s_{p}a_{p}\right|\left|\Phi\left(t\right)\right|^{2}e^{-i\omega_{0}t\left[\hat{\sigma}_{z}^{\left(S\right)}+\hat{\sigma}_{z}^{\left(A\right)}\right]}e^{\hat{\sigma}_{z}^{\left(A\right)}\underset{k}{\sum}\left[g_{k}\hat{b}_{k}^{\dagger}\varphi_{k}\left(t\right)-g_{k}^{*}\hat{b}_{k}\varphi_{k}^{*}\left(t\right)\right]}\\
 & \times & \hat{\rho}_{S}\left(0\right)\hat{\rho}_{B}\left(0\right)e^{i\omega_{0}t\left[\hat{\sigma}_{z}^{\left(S\right)}+\hat{\sigma}_{z}^{\left(A\right)}\right]}e^{-\hat{\sigma}_{z}^{\left(A\right)}\underset{k}{\sum}\left[g_{k}\hat{b}_{k}^{\dagger}\varphi_{k}\left(t\right)-g_{k}^{*}\hat{b}_{k}\varphi_{k}^{*}\left(t\right)\right]}\left|s_{q}a_{q}\right\rangle \\
 & = & \left|\Phi\left(t\right)\right|^{2}e^{-i\omega_{0}t\left(s_{p}+a_{p}\right)}e^{i\omega_{0}t\left(s_{q}+a_{q}\right)}\left\langle s_{p}a_{p}\right|\hat{\rho}_{SA}\left(0\right)\left|s_{q}a_{q}\right\rangle \\
 & \times & e^{a_{p}\underset{k}{\sum}\left[g_{k}\hat{b}_{k}^{\dagger}\varphi_{k}\left(t\right)-g_{k}^{*}\hat{b}_{k}\varphi_{k}^{*}\left(t\right)\right]}\hat{\rho}_{B}\left(0\right)e^{-a_{q}\underset{k}{\sum}\left[g_{k}\hat{b}_{k}^{\dagger}\varphi_{k}\left(t\right)-g_{k}^{*}\hat{b}_{k}\varphi_{k}^{*}\left(t\right)\right]}
\end{eqnarray*}
Then, we take the partial trace over the environmental degrees of
freedom

\begin{eqnarray}
\left\langle s_{p}a_{p}\right|\hat{\rho}_{S}\left(t\right)\left|s_{q}a_{q}\right\rangle  & = & \left\langle s_{p}a_{p}\right|\mathrm{Tr}{}_{B}\left\{ \hat{\rho}_{SB}\left(t\right)\right\} \left|s_{q}a_{q}\right\rangle \nonumber \\
 & = & \left|\Phi\left(t\right)\right|^{2}e^{-i\omega_{0}t\left(s_{p}+a_{p}\right)}e^{i\omega_{0}t\left(s_{q}+a_{q}\right)}\left\langle s_{p}a_{p}\right|\hat{\rho}_{SA}\left(0\right)\left|s_{q}a_{q}\right\rangle \nonumber \\
 & \times & \mathrm{Tr}{}_{B}\left\{ e^{a_{p}\underset{k}{\sum}\left[g_{k}\hat{b}_{k}^{\dagger}\varphi_{k}\left(t\right)-g_{k}^{*}\hat{b}_{k}\varphi_{k}^{*}\left(t\right)\right]}\hat{\rho}_{B}\left(0\right)e^{-a_{q}\underset{k}{\sum}\left[g_{k}\hat{b}_{k}^{\dagger}\varphi_{k}\left(t\right)-g_{k}^{*}\hat{b}_{k}\varphi_{k}^{*}\left(t\right)\right]}\right\} \nonumber \\
\label{rs1}
\end{eqnarray}
We will, for simplicity, treat the term that refers to the environment
separately

\begin{equation}
K\equiv\mathrm{Tr}{}_{B}\left\{ e^{a_{p}\underset{k}{\sum}\left[g_{k}\hat{b}_{k}^{\dagger}\varphi_{k}\left(t\right)-g_{k}^{*}\hat{b}_{k}\varphi_{k}^{*}\left(t\right)\right]}\hat{\rho}_{B}\left(0\right)e^{-a_{q}\underset{k}{\sum}\left[g_{k}\hat{b}_{k}^{\dagger}\varphi_{k}\left(t\right)-g_{k}^{*}\hat{b}_{k}\varphi_{k}^{*}\left(t\right)\right]}\right\} 
\end{equation}
We find that

\begin{eqnarray*}
K & = & \mathrm{Tr}{}_{B}\left\{ \hat{\rho}_{B}\left(0\right)e^{\left(a_{p}-a_{q}\right)\underset{k}{\sum}\left[g_{k}\hat{b}_{k}^{\dagger}\varphi_{k}\left(t\right)-g_{k}^{*}\hat{b}_{k}\varphi_{k}^{*}\left(t\right)\right]}\right\} 
\end{eqnarray*}
where we employed the cyclical property of the trace, $\mathrm{Tr}\left\{ \hat{A}\hat{B}\right\} =\mathrm{Tr}\left\{ \hat{B}\hat{A}\right\} $
for any pair of operators $\hat{A}$ and $\hat{B}$.

Now it is convenient to dissect the exponential, leaving the terms
involving $\hat{b}^{\dagger}$ to the right. Again employing the formula
(\ref{BCH}), with $\begin{cases}
\hat{A} & =-\left(a_{p}-a_{q}\right)\underset{k}{\sum}g_{k}^{*}\hat{b}_{k}\varphi_{k}^{*}\left(t\right)\\
\hat{B} & =\left(a_{p}-a_{q}\right)\underset{k}{\sum}g_{k}\hat{b}_{k}^{\dagger}\varphi_{k}\left(t\right)
\end{cases}$

\begin{eqnarray}
\left[\hat{A},\hat{B}\right] & = & \left[-\left(a_{p}-a_{q}\right)\underset{k}{\sum}g_{k}^{*}\hat{b}_{k}\varphi_{k}^{*}\left(t\right),\left(a_{p}-a_{q}\right)\underset{k}{\sum}g_{k}\hat{b}_{k}^{\dagger}\varphi_{k}\left(t\right)\right]\nonumber \\
 & = & -\left(a_{p}-a_{q}\right)^{2}\underset{k}{\sum}\left|g_{k}\right|^{2}\left|\varphi_{k}\left(t\right)\right|^{2}\label{APBaux1}
\end{eqnarray}
hence

\begin{equation}
e^{\left(a_{p}-a_{q}\right)\underset{k}{\sum}\left[g_{k}\hat{b}_{k}^{\dagger}\varphi_{k}\left(t\right)-g_{k}^{*}\hat{b}_{k}\varphi_{k}^{*}\left(t\right)\right]}=e^{-\left(a_{p}-a_{q}\right)\underset{k}{\sum}g_{k}^{*}\hat{b}_{k}\varphi_{k}^{*}\left(t\right)}e^{\left(a_{p}-a_{q}\right)\underset{k}{\sum}g_{k}\hat{b}_{k}^{\dagger}\varphi_{k}\left(t\right)}e^{\frac{\left(a_{p}-a_{q}\right)^{2}}{2}\underset{k}{\sum}\left|g_{k}\right|^{2}\left|\varphi_{k}\left(t\right)\right|^{2}}
\end{equation}
and
\begin{eqnarray}
K & = & e^{\frac{\left(a_{p}-a_{q}\right)^{2}}{2}\underset{k}{\sum}\left|g_{k}\right|^{2}\left|\varphi_{k}\left(t\right)\right|^{2}}\nonumber \\
 & \times & \mathrm{Tr}{}_{B}\left\{ e^{\left(a_{p}-a_{q}\right)\underset{k}{\sum}g_{k}\hat{b}_{k}^{\dagger}\varphi_{k}\left(t\right)}\underset{k}{\prod}\frac{1}{Z_{k}}\underset{n}{\sum}\left|n\right\rangle \left\langle n\right|\exp\left(-n\beta\hbar\omega_{k}\right)e^{-\left(a_{p}-a_{q}\right)\underset{k}{\sum}g_{k}^{*}\hat{b}_{k}\varphi_{k}^{*}\left(t\right)}\right\} \label{K1}
\end{eqnarray}

As the modes of the bath are independent, (\ref{K1}) can be re-written
as 

\begin{eqnarray}
K & = & e^{\frac{\left(a_{p}-a_{q}\right)^{2}}{2}\underset{k}{\sum}\left|g_{k}\right|^{2}\left|\varphi_{k}\left(t\right)\right|^{2}}\nonumber \\
 & \times & \underset{k}{\prod}\frac{1}{Z_{k}}\underset{n}{\sum}\mathrm{Tr}{}_{B_{k}}\left\{ e^{\left(a_{p}-a_{q}\right)g_{k}\hat{b}_{k}^{\dagger}\varphi_{k}\left(t\right)}\left|n\right\rangle \left\langle n\right|\exp\left(-n\beta\hbar\omega_{k}\right)e^{-\left(a_{p}-a_{q}\right)g_{k}^{*}\hat{b}_{k}\varphi_{k}^{*}\left(t\right)}\right\} \label{K2}
\end{eqnarray}

Let us calculate the trace directly from the basis of coherent states:
$\left|\alpha_{k}\right\rangle =e^{-\frac{\left|\alpha_{k}\right|^{2}}{2}}\underset{m}{\sum}\frac{\alpha_{k}^{m}}{\sqrt{m!}}\left|m\right\rangle $
\begin{eqnarray}
 & \mathrm{Tr}{}_{B_{k}}\left\{ e^{\left(a_{p}-a_{q}\right)g_{k}\hat{b}_{k}^{\dagger}\varphi_{k}\left(t\right)}\left|n\right\rangle \left\langle n\right|\exp\left(-n\beta\hbar\omega_{k}\right)e^{-\left(a_{p}-a_{q}\right)g_{k}^{*}\hat{b}_{k}\varphi_{k}^{*}\left(t\right)}\right\} \nonumber \\
= & \int\frac{d^{2}\alpha_{k}}{\pi}\left\langle \alpha_{k}\right|e^{\left(a_{p}-a_{q}\right)g_{k}\hat{b}_{k}^{\dagger}\varphi_{k}\left(t\right)}\left|n\right\rangle \left\langle n\right|\exp\left(-n\beta\hbar\omega_{k}\right)e^{-\left(a_{p}-a_{q}\right)g_{k}^{*}\hat{b}_{k}\varphi_{k}^{*}\left(t\right)}\left|\alpha_{k}\right\rangle \nonumber \\
= & \int\frac{d^{2}\alpha_{k}}{\pi}e^{\left(a_{p}-a_{q}\right)g_{k}\varphi_{k}\left(t\right)\alpha_{k}^{*}}\left\langle \alpha_{k}|n\right\rangle \left\langle n|\alpha_{k}\right\rangle \exp\left(-n\beta\hbar\omega_{k}\right)e^{-\left(a_{p}-a_{q}\right)g_{k}^{*}\alpha_{k}\varphi_{k}^{*}\left(t\right)}\nonumber \\
= & \int\frac{d^{2}\alpha_{k}}{\pi}e^{\left(a_{p}-a_{q}\right)g_{k}\varphi_{k}\left(t\right)\alpha_{k}^{*}}\left|\left\langle n|\alpha_{k}\right\rangle \right|^{2}\exp\left(-n\beta\hbar\omega_{k}\right)e^{-\left(a_{p}-a_{q}\right)g_{k}^{*}\alpha_{k}\varphi_{k}^{*}\left(t\right)}
\end{eqnarray}
and, as

\begin{equation}
\left\langle n|\alpha_{k}\right\rangle =\left\langle n\right|e^{-\frac{\left|\alpha_{k}\right|^{2}}{2}}\underset{m}{\sum}\frac{\alpha_{k}^{m}}{\sqrt{m!}}\left|m\right\rangle =e^{-\frac{\left|\alpha_{k}\right|^{2}}{2}}\underset{m}{\sum}\frac{\alpha_{k}^{m}}{\sqrt{m!}}\delta_{m,n}=e^{-\frac{\left|\alpha_{k}\right|^{2}}{2}}\frac{\alpha_{k}^{n}}{\sqrt{n!}}
\end{equation}
we have

\begin{eqnarray}
 & \mathrm{Tr}{}_{B_{k}}\left\{ e^{\left(a_{p}-a_{q}\right)g_{k}\hat{b}_{k}^{\dagger}\varphi_{k}\left(t\right)}\left|n\right\rangle \left\langle n\right|\exp\left(-n\beta\hbar\omega_{k}\right)e^{-\left(a_{p}-a_{q}\right)g_{k}^{*}\hat{b}_{k}\varphi_{k}^{*}\left(t\right)}\right\} \nonumber \\
= & \int\frac{d^{2}\alpha_{k}}{\pi}e^{\left(a_{p}-a_{q}\right)g_{k}\varphi_{k}\left(t\right)\alpha_{k}^{*}}e^{-\left|\alpha_{k}\right|^{2}}\frac{\left|\alpha_{k}\right|^{2n}}{n!}\exp\left(-n\beta\hbar\omega_{k}\right)e^{-\left(a_{p}-a_{q}\right)g_{k}^{*}\alpha_{k}\varphi_{k}^{*}\left(t\right)}\nonumber \\
= & \int\frac{d^{2}\alpha_{k}}{\pi}e^{\left(a_{p}-a_{q}\right)\left[g_{k}\varphi_{k}\left(t\right)\alpha_{k}^{*}-g_{k}^{*}\alpha_{k}\varphi_{k}^{*}\left(t\right)\right]}e^{-\left|\alpha_{k}\right|^{2}}\frac{\left|\alpha_{k}\right|^{2n}}{n!}\exp\left(-n\beta\hbar\omega_{k}\right)
\end{eqnarray}

Back to (\ref{K2}),

\begin{eqnarray}
K & = & e^{\frac{\left(a_{p}-a_{q}\right)^{2}}{2}\underset{k}{\sum}\left|g_{k}\right|^{2}\left|\varphi_{k}\left(t\right)\right|^{2}}\underset{k}{\prod}\frac{1}{Z_{k}}\int\frac{d^{2}\alpha_{k}}{\pi}e^{\left(a_{p}-a_{q}\right)\left[g_{k}\varphi_{k}\left(t\right)\alpha_{k}^{*}-g_{k}^{*}\alpha_{k}\varphi_{k}^{*}\left(t\right)\right]}e^{-\left|\alpha_{k}\right|^{2}\left(1-e^{-\beta\hbar\omega_{k}}\right)}\nonumber \\
\label{K3}
\end{eqnarray}

For convenience, let us reproduce some definitions from (\ref{K3}):

\begin{equation}
\begin{cases}
\alpha_{k} & =x+iy\\
g_{k}\varphi_{k}\left(t\right) & =A+iB\\
\left(1-e^{-\beta\hbar\omega_{k}}\right) & =C
\end{cases}\label{APBaux2}
\end{equation}
(do not mistake $A$ and $B$ in (\ref{APBaux2}) with the operators
$\hat{A}$ and $\hat{B}$ in (\ref{APBaux1}) or with the measurement
apparatus $A$ and the environment $B$). Thus, $\mathrm{d}^{2}\alpha_{k}=\mathrm{d}x\mathrm{d}y$
and

\begin{eqnarray}
K & = & e^{\frac{\left(a_{p}-a_{q}\right)^{2}}{2}\underset{k}{\sum}\left|g_{k}\right|^{2}\left|\varphi_{k}\left(t\right)\right|^{2}}\underset{k}{\prod}\frac{1}{\pi Z_{k}}\int_{-\infty}^{\infty}\mathrm{d}x\int_{-\infty}^{\infty}\mathrm{d}ye^{\left(a_{p}-a_{q}\right)\left[\left(A+iB\right)\left(x-iy\right)-\left(A-iB\right)\left(x+iy\right)\right]}e^{-\left(x^{2}+y^{2}\right)C}\nonumber \\
 & = & e^{\frac{\left(a_{p}-a_{q}\right)^{2}}{2}\underset{k}{\sum}\left|g_{k}\right|^{2}\left|\varphi_{k}\left(t\right)\right|^{2}}\underset{k}{\prod}\frac{1}{\pi Z_{k}}\int_{-\infty}^{\infty}\mathrm{d}x\int_{-\infty}^{\infty}\mathrm{d}ye^{\left(a_{p}-a_{q}\right)\left[Ax-iAy+iBx+By-\left(Ax+iAy-iBx+By\right)\right]}e^{-\left(x^{2}+y^{2}\right)C}\nonumber \\
 & = & e^{\frac{\left(a_{p}-a_{q}\right)^{2}}{2}\underset{k}{\sum}\left|g_{k}\right|^{2}\left|\varphi_{k}\left(t\right)\right|^{2}}\underset{k}{\prod}\frac{1}{\pi Z_{k}}\int_{-\infty}^{\infty}\mathrm{d}x\int_{-\infty}^{\infty}\mathrm{d}ye^{i2\left(a_{p}-a_{q}\right)\left(Bx-Ay\right)}e^{-\left(x^{2}+y^{2}\right)C}\nonumber \\
 & = & e^{\frac{\left(a_{p}-a_{q}\right)^{2}}{2}\underset{k}{\sum}\left|g_{k}\right|^{2}\left|\varphi_{k}\left(t\right)\right|^{2}}\underset{k}{\prod}\frac{1}{\pi Z_{k}}\int_{-\infty}^{\infty}\mathrm{d}xe^{i2\left(a_{p}-a_{q}\right)Bx}e^{-Cx^{2}}\int_{-\infty}^{\infty}\mathrm{d}ye^{-i2\left(a_{p}-a_{q}\right)Ay}e^{-Cy^{2}}\nonumber \\
\end{eqnarray}

Each integral yields:

\begin{eqnarray}
\int_{-\infty}^{\infty}\mathrm{d}xe^{i2\left(a_{p}-a_{q}\right)Bx}e^{-Cx^{2}} & = & \int_{-\infty}^{\infty}\mathrm{d}xe^{-Cx^{2}}\cos\left[2\left(a_{p}-a_{q}\right)Bx\right]\nonumber \\
 & + & i\int_{-\infty}^{\infty}\mathrm{d}xe^{-Cx^{2}}\sin\left[2\left(a_{p}-a_{q}\right)Bx\right]\nonumber \\
\nonumber \\
\int_{-\infty}^{\infty}\mathrm{d}ye^{-i2\left(a_{p}-a_{q}\right)Ay}e^{-Cy^{2}} & = & \int_{-\infty}^{\infty}\mathrm{d}ye^{-Cy^{2}}\cos\left[i2\left(a_{p}-a_{q}\right)Ay\right]\nonumber \\
 & - & i\int_{-\infty}^{\infty}\mathrm{d}ye^{-Cy^{2}}\sin\left[i2\left(a_{p}-a_{q}\right)Ay\right]
\end{eqnarray}
The integrals of the imaginary parts will vanish (because they are
odd integrands over symmetric intervals). Hence,
\begin{eqnarray}
\int_{-\infty}^{\infty}\mathrm{d}xe^{i2\left(a_{p}-a_{q}\right)Bx}e^{-Cx^{2}} & = & \int_{-\infty}^{\infty}\mathrm{d}xe^{-Cx^{2}}\cos\left[2\left(a_{p}-a_{q}\right)Bx\right]\nonumber \\
\int_{-\infty}^{\infty}\mathrm{d}ye^{-i2\left(a_{p}-a_{q}\right)Ay}e^{-Cy^{2}} & = & \int_{-\infty}^{\infty}\mathrm{d}ye^{-Cy^{2}}\cos\left[i2\left(a_{p}-a_{q}\right)Ay\right]
\end{eqnarray}
Employing the result \cite{key-57,key-58}

\begin{equation}
\int_{-\infty}^{\infty}\mathrm{d}xe^{-ax^{2}}\cos\left(bx\right)=\sqrt{\frac{\pi}{a}}e^{-\frac{b^{2}}{4a}}
\end{equation}
we find
\begin{eqnarray}
\int_{-\infty}^{\infty}\mathrm{d}xe^{i2\left(a_{p}-a_{q}\right)Bx}e^{-Cx^{2}} & = & \sqrt{\frac{\pi}{C}}e^{-\frac{B^{2}}{C}\left(a_{p}-a_{q}\right)^{2}}\nonumber \\
\int_{-\infty}^{\infty}\mathrm{d}ye^{-i2\left(a_{p}-a_{q}\right)Ay}e^{-Cy^{2}} & = & \sqrt{\frac{\pi}{C}}e^{-\frac{A^{2}}{C}\left(a_{p}-a_{q}\right)^{2}}
\end{eqnarray}
or

\begin{eqnarray}
K & = & e^{\frac{\left(a_{p}-a_{q}\right)^{2}}{2}\underset{k}{\sum}\left|g_{k}\right|^{2}\left|\varphi_{k}\left(t\right)\right|^{2}}\underset{k}{\prod}\frac{1}{\pi Z_{k}}\sqrt{\frac{\pi}{C}}e^{-\frac{B^{2}}{C}\left(a_{p}-a_{q}\right)^{2}}\sqrt{\frac{\pi}{C}}e^{-\frac{A^{2}}{C}\left(a_{p}-a_{q}\right)^{2}}\nonumber \\
 & = & e^{\frac{\left(a_{p}-a_{q}\right)^{2}}{2}\underset{k}{\sum}\left|g_{k}\right|^{2}\left|\varphi_{k}\left(t\right)\right|^{2}}\underset{k}{\prod}\frac{1}{CZ_{k}}e^{-\frac{A^{2}+B^{2}}{C}\left(a_{p}-a_{q}\right)^{2}}
\end{eqnarray}

As

\begin{equation}
A^{2}+B^{2}=\left|g_{k}\right|^{2}\left|\varphi_{k}\left(t\right)\right|^{2}
\end{equation}
then

\begin{equation}
K=e^{\frac{\left(a_{p}-a_{q}\right)^{2}}{2}\underset{k}{\sum}\left|g_{k}\right|^{2}\left|\varphi_{k}\left(t\right)\right|^{2}}\underset{k}{\prod}\frac{1}{\left(1-e^{-\beta\hbar\omega_{k}}\right)Z_{k}}e^{-\frac{\left|g_{k}\right|^{2}\left|\varphi_{k}\left(t\right)\right|^{2}}{1-e^{-\beta\hbar\omega_{k}}}\left(a_{p}-a_{q}\right)^{2}}
\end{equation}
and, from the definition of $Z_{k}$,
\begin{eqnarray}
K & = & e^{\frac{\left(a_{p}-a_{q}\right)^{2}}{2}\underset{k}{\sum}\left|g_{k}\right|^{2}\left|\varphi_{k}\left(t\right)\right|^{2}}\underset{k}{\prod}e^{-\frac{\left|g_{k}\right|^{2}\left|\varphi_{k}\left(t\right)\right|^{2}}{1-e^{-\beta\hbar\omega_{k}}}\left(a_{p}-a_{q}\right)^{2}}\nonumber \\
 & = & e^{\left(a_{p}-a_{q}\right)^{2}\underset{k}{\sum}\left|g_{k}\right|^{2}\left|\varphi_{k}\left(t\right)\right|^{2}\left[\frac{1}{2}-\frac{1}{1-e^{-\beta\hbar\omega_{k}}}\right]}
\end{eqnarray}
but

\begin{equation}
\frac{1}{2}-\frac{1}{1-e^{-\beta\hbar\omega_{k}}}=\frac{1-e^{-\beta\hbar\omega_{k}}-2}{2\left(1-e^{-\beta\hbar\omega_{k}}\right)}=-\frac{1}{2}\frac{1+e^{-\beta\hbar\omega_{k}}}{1-e^{-\beta\hbar\omega_{k}}}=-\frac{1}{2}\coth\left(\frac{\beta\hbar\omega_{k}}{2}\right)
\end{equation}
therefore

\begin{equation}
K=e^{-\frac{\left(a_{p}-a_{q}\right)}{2}^{2}\underset{k}{\sum}\left|g_{k}\right|^{2}\left|\varphi_{k}\left(t\right)\right|^{2}\coth\left(\frac{\beta\hbar\omega_{k}}{2}\right)}
\end{equation}

Returning to (\ref{rs1}), we have, finally,
\begin{eqnarray}
\left\langle s_{p}a_{p}\right|\hat{\rho}_{SA}\left(t\right)\left|s_{q}a_{q}\right\rangle  & = & \left|\Phi\left(t\right)\right|^{2}e^{-i\omega_{0}t\left(s_{p}+a_{p}\right)}e^{i\omega_{0}t\left(s_{q}+a_{q}\right)}\nonumber \\
 & \times & \left\langle s_{p}a_{p}\right|\hat{\rho}_{SA}\left(0\right)\left|s_{q}a_{q}\right\rangle e^{-\frac{\left(a_{p}-a_{q}\right)}{2}^{2}\underset{k}{\sum}\left|g_{k}\right|^{2}\left|\varphi_{k}\left(t\right)\right|^{2}\coth\left(\frac{\beta\hbar\omega_{k}}{2}\right)}.\label{Exp.Apendice}
\end{eqnarray}

\section*{Appendix C: The Continuous Limit}

Now that the partial trace has been take, we can proceed to the limit
to the continuum. In the first place, we define a spectral density

\begin{equation}
J\left(\omega\right)=\underset{k}{\sum}\left|g_{k}\right|^{2}\delta\left(\omega-\omega_{k}\right)
\end{equation}
so that (\ref{Exp.Apendice}) becomes

\begin{eqnarray}
\left\langle s_{p}a_{p}\right|\hat{\rho}_{S}\left(t\right)\left|s_{q}a_{q}\right\rangle  & = & \left|\Phi\left(t\right)\right|^{2}e^{-i\omega_{0}t\left(s_{p}-s_{q}\right)}e^{-i\omega_{0}t\left(a_{p}-a_{q}\right)}\nonumber \\
 & \times & e^{-\frac{\left(a_{p}-a_{q}\right)}{2}^{2}\int_{0}^{\infty}d\omega J\left(\omega\right)\left|\varphi\left(t\right)\right|^{2}\coth\left(\frac{\beta\hbar\omega}{2}\right)}\left\langle s_{p}a_{p}\right|\hat{\rho}_{S}\left(0\right)\left|s_{q}a_{q}\right\rangle \label{rs2}
\end{eqnarray}

\begin{equation}
\varphi\left(t\right)=\frac{1-e^{i\omega t}}{\omega}
\end{equation}
and
\begin{equation}
\Phi\left(t\right)=e^{-\int_{0}^{\infty}\mathrm{d}\omega J\left(\omega\right)\frac{\varphi^{*}\left(t\right)}{\omega}}e^{it\int_{0}^{\infty}\mathrm{d}\omega\frac{J\left(\omega\right)}{\omega}}e^{\frac{1}{2}\int_{0}^{\infty}\mathrm{d}\omega J\left(\omega\right)\left|\varphi\left(t\right)\right|^{2}}
\end{equation}

Let us analyse $\left|\varphi\left(t\right)\right|^{2}$ and $\left|\Phi\left(t\right)\right|^{2}$:

\begin{eqnarray}
\left|\varphi\left(t\right)\right|^{2} & = & \left|\frac{1-e^{i\omega t}}{\omega}\right|^{2}=\frac{1-e^{i\omega t}}{\omega}\frac{1-e^{-i\omega t}}{\omega}\nonumber \\
 & = & \frac{1-e^{-i\omega t}-e^{i\omega t}+1}{\omega^{2}}=\frac{2-2\cos\left(\omega t\right)}{\omega^{2}}\nonumber \\
 & = & 2\frac{1-\cos\left(\omega t\right)}{\omega^{2}}
\end{eqnarray}

\begin{eqnarray*}
\left|\Phi\left(t\right)\right|^{2} & = & \Phi\left(t\right)\Phi^{*}\left(t\right)\\
 & = & e^{-\int_{0}^{\infty}\mathrm{d}\omega J\left(\omega\right)\frac{\varphi^{*}\left(t\right)}{\omega}}e^{it\int_{0}^{\infty}\mathrm{d}\omega\frac{J\left(\omega\right)}{\omega}}e^{\frac{1}{2}\int_{0}^{\infty}\mathrm{d}\omega J\left(\omega\right)\left|\varphi\left(t\right)\right|^{2}}\\
 & \times & e^{-\int_{0}^{\infty}\mathrm{d}\omega J\left(\omega\right)\frac{\varphi\left(t\right)}{\omega}}e^{-it\int_{0}^{\infty}\mathrm{d}\omega\frac{J\left(\omega\right)}{\omega}}e^{\frac{1}{2}\int_{0}^{\infty}\mathrm{d}\omega J\left(\omega\right)\left|\varphi\left(t\right)\right|^{2}}\\
 & = & e^{-\int_{0}^{\infty}\mathrm{d}\omega J\left(\omega\right)\frac{\varphi\left(t\right)+\varphi^{*}\left(t\right)}{\omega}}e^{\int_{0}^{\infty}\mathrm{d}\omega J\left(\omega\right)\left|\varphi\left(t\right)\right|^{2}}\\
 & = & e^{-\int_{0}^{\infty}\mathrm{d}\omega J\left(\omega\right)\frac{2Re\left\{ \varphi\left(t\right)\right\} }{\omega}}e^{\int_{0}^{\infty}\mathrm{d}\omega J\left(\omega\right)\left|\varphi\left(t\right)\right|^{2}}\\
 & = & e^{-2\int_{0}^{\infty}\mathrm{d}\omega J\left(\omega\right)\frac{1-cos\left(\omega t\right)}{\omega^{2}}}e^{2\int_{0}^{\infty}\mathrm{d}\omega J\left(\omega\right)\frac{1-cos\left(\omega t\right)}{\omega^{2}}}\\
 & = & 1
\end{eqnarray*}
Returning to (\ref{rs2}):

\begin{eqnarray}
\left\langle s_{p}a_{p}\right|\hat{\rho}_{S}\left(t\right)\left|s_{q}a_{q}\right\rangle  & = & e^{-i\omega_{0}t\left(s_{p}-s_{q}\right)}e^{-i\omega_{0}t\left(a_{p}-a_{q}\right)}\nonumber \\
 & \times & e^{-\left(a_{p}-a_{q}\right)^{2}\int_{0}^{\infty}d\omega J\left(\omega\right)\frac{1-\cos\left(\omega t\right)}{\omega^{2}}\coth\left(\frac{\beta\hbar\omega}{2}\right)}\left\langle s_{p}a_{p}\right|\hat{\rho}_{S}\left(0\right)\left|s_{q}a_{q}\right\rangle \label{rs3-1}
\end{eqnarray}

Finding the matrix elements depends, therefore, on determining the
integral

\begin{equation}
I_{1}\left(t\right)=\int_{0}^{\infty}d\omega J\left(\omega\right)\frac{1-\cos\left(\omega t\right)}{\omega^{2}}\coth\left(\frac{\beta\hbar\omega}{2}\right)\label{integrais-1}
\end{equation}
Re-writing (\ref{rs3-1}), we find the final formula

\begin{equation}
\left\langle s_{p}a_{p}\right|\hat{\rho}_{S}\left(t\right)\left|s_{q}a_{q}\right\rangle =e^{-i\omega_{0}t\left(s_{p}-s_{q}\right)}e^{-i\omega_{0}t\left(a_{p}-a_{q}\right)}e^{-\left(a_{p}-a_{q}\right)^{2}I_{1}\left(t\right)}\left\langle s_{p}a_{p}\right|\hat{\rho}_{S}\left(0\right)\left|s_{q}a_{q}\right\rangle \label{rs4-1}
\end{equation}

As the density matrix is Hermitian, we just to determine 10 out of
16 matrix elements. For a general state

\begin{equation}
\hat{\rho}_{S}\left(0\right)=\left(\begin{array}{cccc}
\rho_{11} & \rho_{12} & \rho_{13} & \rho_{14}\\
\rho_{12}^{*} & \rho_{22} & \rho_{23} & \rho_{24}\\
\rho_{13}^{*} & \rho_{23}^{*} & \rho_{33} & \rho_{34}\\
\rho_{14}^{*} & \rho_{24}^{*} & \rho_{34}^{*} & \rho_{44}
\end{array}\right)
\end{equation}

\begin{itemize}
\item $\left\langle ++\left|\hat{\rho}_{S}\left(t\right)\right|++\right\rangle $:$\begin{cases}
s_{p}= & 1\\
s_{q}= & 1\\
a_{p}= & 1\\
a_{q}= & 1
\end{cases}$
\end{itemize}
\begin{equation}
\left\langle ++\right|\hat{\rho}_{S}\left(t\right)\left|++\right\rangle =\rho_{11}
\end{equation}

\begin{itemize}
\item $\left\langle ++\left|\hat{\rho}_{S}\left(t\right)\right|+-\right\rangle $:$\begin{cases}
s_{p}= & 1\\
s_{q}= & 1\\
a_{p}= & 1\\
a_{q}= & -1
\end{cases}$
\end{itemize}
\begin{equation}
\left\langle ++\right|\hat{\rho}_{S}\left(t\right)\left|+-\right\rangle =e^{-i2\omega_{0}t}e^{-4I_{1}\left(t\right)}\left\langle ++\right|\hat{\rho}_{S}\left(0\right)\left|+-\right\rangle =e^{-i2\omega_{0}t}e^{-4I_{1}\left(t\right)}\rho_{12}
\end{equation}

\begin{itemize}
\item $\left\langle ++\left|\hat{\rho}_{S}\left(t\right)\right|-+\right\rangle $:$\begin{cases}
s_{p}= & 1\\
s_{q}= & -1\\
a_{p}= & 1\\
a_{q}= & 1
\end{cases}$
\end{itemize}
\begin{equation}
\left\langle ++\right|\hat{\rho}_{S}\left(t\right)\left|-+\right\rangle =e^{-i2\omega_{0}t}\left\langle ++\right|\hat{\rho}_{S}\left(0\right)\left|-+\right\rangle =e^{-i2\omega_{0}t}\rho_{13}
\end{equation}

\begin{itemize}
\item $\left\langle ++\left|\hat{\rho}_{S}\left(t\right)\right|--\right\rangle $:$\begin{cases}
s_{p}= & 1\\
s_{q}= & -1\\
a_{p}= & 1\\
a_{q}= & -1
\end{cases}$
\end{itemize}
\begin{equation}
\left\langle ++\right|\hat{\rho}_{S}\left(t\right)\left|--\right\rangle =e^{-i2\omega_{0}t}e^{-i2\omega_{0}t}e^{-4I_{1}\left(t\right)}\left\langle ++\right|\hat{\rho}_{S}\left(0\right)\left|--\right\rangle =e^{-i4\omega_{0}t}e^{-4I_{1}\left(t\right)}\rho_{14}
\end{equation}

\begin{itemize}
\item $\left\langle +-\left|\hat{\rho}_{S}\left(t\right)\right|+-\right\rangle $:$\begin{cases}
s_{p}= & 1\\
s_{q}= & 1\\
a_{p}= & -1\\
a_{q}= & -1
\end{cases}$
\end{itemize}
\begin{equation}
\left\langle +-\right|\hat{\rho}_{S}\left(t\right)\left|+-\right\rangle =\left\langle +-\right|\hat{\rho}_{S}\left(0\right)\left|+-\right\rangle =\rho_{22}
\end{equation}

\begin{itemize}
\item $\left\langle +-\left|\hat{\rho}_{S}\left(t\right)\right|-+\right\rangle $:$\begin{cases}
s_{p}= & 1\\
s_{q}= & -1\\
a_{p}= & -1\\
a_{q}= & 1
\end{cases}$
\end{itemize}
\begin{equation}
\left\langle +-\right|\hat{\rho}_{S}\left(t\right)\left|-+\right\rangle =e^{-i2\omega_{0}t}e^{i2\omega_{0}t}e^{-4I_{1}\left(t\right)}\left\langle +-\right|\hat{\rho}_{S}\left(0\right)\left|-+\right\rangle =e^{-4I_{1}\left(t\right)}\rho_{23}
\end{equation}

\begin{itemize}
\item $\left\langle +-\left|\hat{\rho}_{S}\left(t\right)\right|--\right\rangle $:$\begin{cases}
s_{p}= & 1\\
s_{q}= & -1\\
a_{p}= & -1\\
a_{q}= & -1
\end{cases}$
\end{itemize}
\begin{equation}
\left\langle +-\right|\hat{\rho}_{S}\left(t\right)\left|--\right\rangle =e^{-i2\omega_{0}t}\left\langle +-\right|\hat{\rho}_{S}\left(0\right)\left|--\right\rangle =e^{-i2\omega_{0}t}\rho_{24}
\end{equation}

\begin{itemize}
\item $\left\langle -+\left|\hat{\rho}_{S}\left(t\right)\right|-+\right\rangle $:$\begin{cases}
s_{p}= & -1\\
s_{q}= & -1\\
a_{p}= & 1\\
a_{q}= & 1
\end{cases}$
\end{itemize}
\begin{equation}
\left\langle -+\right|\hat{\rho}_{S}\left(t\right)\left|-+\right\rangle =\rho_{33}
\end{equation}

\begin{itemize}
\item $\left\langle -+\left|\hat{\rho}_{S}\left(t\right)\right|--\right\rangle $:$\begin{cases}
s_{p}= & -1\\
s_{q}= & -1\\
a_{p}= & 1\\
a_{q}= & -1
\end{cases}$
\end{itemize}
\begin{equation}
\left\langle -+\right|\hat{\rho}_{S}\left(t\right)\left|--\right\rangle =e^{-i2\omega_{0}t}e^{-4I_{1}\left(t\right)}\left\langle -+\right|\hat{\rho}_{S}\left(0\right)\left|--\right\rangle =e^{-i2\omega_{0}t}e^{-4I_{1}\left(t\right)}\rho_{34}
\end{equation}

\begin{itemize}
\item $\left\langle --\left|\hat{\rho}_{S}\left(t\right)\right|--\right\rangle $:$\begin{cases}
s_{p}= & -1\\
s_{q}= & -1\\
a_{p}= & -1\\
a_{q}= & -1
\end{cases}$
\end{itemize}
\begin{equation}
\left\langle --\right|\hat{\rho}_{S}\left(t\right)\left|--\right\rangle =\left\langle --\right|\hat{\rho}_{S}\left(0\right)\left|--\right\rangle =\rho_{44}
\end{equation}

In the end,

\begin{equation}
\hat{\rho}_{S}\left(t\right)=\left(\begin{array}{cccc}
\rho_{11} & e^{-i2\omega_{0}t}e^{-4I_{1}\left(t\right)}\rho_{12} & e^{-i2\omega_{0}t}\rho_{13} & e^{-i4\omega_{0}t}e^{-4I_{1}\left(t\right)}\rho_{14}\\
e^{i2\omega_{0}t}e^{-4I_{1}\left(t\right)}\rho_{12}^{*} & \rho_{22} & e^{-4I_{1}\left(t\right)}\rho_{23} & e^{-i2\omega_{0}t}\rho_{24}\\
e^{i2\omega_{0}t}\rho_{13}^{*} & e^{-4I_{1}\left(t\right)}\rho_{23}^{*} & \rho_{33} & e^{-i2\omega_{0}t}e^{-4I_{1}\left(t\right)}\rho_{34}\\
e^{i4\omega_{0}t}e^{-4I_{1}\left(t\right)}\rho_{14}^{*} & e^{i2\omega_{0}t}\rho_{24}^{*} & e^{i2\omega_{0}t}e^{-4I_{1}\left(t\right)}\rho_{34}^{*} & \rho_{44}
\end{array}\right)\label{roSt-1}
\end{equation}

We stress that in the limit $t\rightarrow0$, the initial condition
is re-obtained.

\section*{Appendix D: Envariance and Born's rule}

The decoherence program did not end with the explanation of pointer
basis. In its ambition, it went further, trying to shed light into
such fundamental concepts of quantum mechanics as Born's rule.

To explain how Born's rule can be obtained from decoherence, a procedure
that was presented by Zurek in \cite{BORNrule01}, we must first introduce
the concept of \emph{envariance}. A joint state $\left|\psi\right\rangle _{S+A}$
of a principal system $S$ and an environment $A$ is said to be \emph{envariant
}with respect to some transformation \textbf{$U_{S}$} that acts on
the principal system alone if there exists some transformation $U_{A}$
that, acting solely on the environment, is capable of reversing the
action of $U_{S}$:

\[
\left|\psi\right\rangle _{S+A}=U_{A}U_{S}\left|\psi\right\rangle _{S+A}.
\]
Being envariant, therefore, means that the transformation $U_{S}$
is imperceptible if you only have access to the principal system.

It is always possible to decompose the joint state of two Hilbert
spaces into its Schmidt form, that is, into the form

\[
\left|\psi\right\rangle _{S+A}=\underset{i}{\sum}c_{i}\left|s_{i}\right\rangle _{S}\left|a_{i}\right\rangle _{A}
\]
where the $\left\{ \left|s_{i}\right\rangle _{S}\right\} $ and $\left\{ \left|a_{i}\right\rangle _{A}\right\} $
are both orthonormal bases of $S$ and $A$, respectively, chosen
depending on the original joint state $\left|\psi\right\rangle _{S+A}$.

In this article, we are considering a two-level quantum system as
our principal system. In this case, the Schmidt basis will consist
at most of two states:

\[
\left|\psi\right\rangle _{S+A}=c_{0}\left|s_{0}\right\rangle _{S}\left|a_{0}\right\rangle _{A}+c_{1}\left|s_{1}\right\rangle _{S}\left|a_{1}\right\rangle _{A}.
\]
Or, leaving the absolute values of the coefficients $c_{0}$ and $c_{1}$
more explicit,

\[
\left|\psi\right\rangle _{S+A}=\left|c_{0}\right|e^{i\phi_{0}}\left|s_{0}\right\rangle _{S}\left|a_{0}\right\rangle _{A}+\left|c_{1}\right|e^{i\phi_{1}}\left|s_{1}\right\rangle _{S}\left|a_{1}\right\rangle _{A}.
\]

Following Zurek, we introduce the swap transformation for the principal
system:

\[
U_{S}=e^{i\varphi}\left|s_{1}\right\rangle _{S}\left\langle s_{0}\right|_{S}+e^{-i\varphi}\left|s_{0}\right\rangle _{S}\left\langle s_{1}\right|_{S},
\]
which is clearly unitary. This transformation, besides adding a phase,
which, as we will see, is irrelevant to the problem, simply swaps
the basis states of the environment. The joint state $\left|\psi\right\rangle _{S+A}$
will be envariant with respect to $U_{S}$ if $\left|c_{0}\right|=\left|c_{1}\right|$,
because, in this case, we simply apply

\[
U_{A}=e^{i\left(\phi_{1}-\phi_{0}-\varphi\right)}\left|a_{1}\right\rangle _{A}\left\langle a_{0}\right|_{A}+e^{-i\left(\phi_{1}-\phi_{0}-\varphi\right)}\left|a_{0}\right\rangle _{A}\left\langle a_{1}\right|_{A}
\]
to return to the original state.

As we can see, swapping the states $\left|s_{0}\right\rangle _{S}$
and $\left|s_{1}\right\rangle _{S}$ is an action that cannot be noticed
by an observer who has no access to the environment, if $\left|c_{0}\right|=\left|c_{1}\right|$.
We might as well change their labels, and we still would not perceive
any difference. Therefore, the probabilities of measuring $\left|s_{0}\right\rangle _{S}$
or $\left|s_{1}\right\rangle _{S}$ must be the same. This is not
Born's rule, but it is a beginning.

To derive the full Born's rule, we must include an auxiliary system
between our object system and the environment. An observer or measurement
apparatus, if you will. We shall refer to it as $M$.

Suppose that the system-object has only interacted with $M$ so far.
In this case, the Schmidt decomposition can be written as:

\[
\left|\psi\right\rangle _{S+A}=\left(\left|c_{0}\right|e^{i\phi_{0}}\left|s_{0}\right\rangle _{S}\left|m_{0}\right\rangle _{M}+\left|c_{1}\right|e^{i\phi_{1}}\left|s_{1}\right\rangle _{S}\left|m_{1}\right\rangle _{M}\right)\left|a_{0}\right\rangle _{A}.
\]
The observer $M$, however, has many more degrees of freedom than
the object-system $S$. As $\left|m_{0}\right\rangle _{M}$ and $\left|m_{1}\right\rangle _{M}$
are orthogonal (according to the Schmidt decomposition), we will re-write
them as equal-weighed superpositions of some new orthonormal basis
$\left\{ \left|\tilde{m}_{k}\right\rangle _{M}\right\} $:

\[
\left|m_{0}\right\rangle _{M}=\frac{1}{\sqrt{a}}\sum_{k=1}^{a}\left|\tilde{m}_{k}\right\rangle _{M},
\]

\[
\left|m_{1}\right\rangle _{M}=\frac{1}{\sqrt{b}}\sum_{k=a+1}^{a+b}\left|\tilde{m}_{k}\right\rangle _{M}.
\]
where $a$ and $b$ are the degrees of freedom necessary to express
the Schmidt-basis states $\left|m_{0}\right\rangle _{M}$ and $\left|m_{1}\right\rangle _{M}$
, respectively.

Letting the rest of the environment interact with the system-object
and the auxiliary ``observer'', we will reach a final state with
the following Schmidt form:

\[
\left|\psi\right\rangle _{S+A}=\frac{\left|c_{0}\right|}{\sqrt{a}}e^{i\phi_{0}}\sum_{k=1}^{a}\left|s_{0}\right\rangle _{S}\left|\tilde{m}_{k}\right\rangle _{M}\left|\tilde{a}_{k}\right\rangle _{A}+\frac{\left|c_{1}\right|}{\sqrt{b}}e^{i\phi_{1}}\sum_{k=a+1}^{b}\left|s_{1}\right\rangle _{S}\left|\tilde{m}_{k}\right\rangle _{M}\left|\tilde{a}_{k}\right\rangle _{A},
\]
where $\left\{ \left|\tilde{a}_{k}\right\rangle _{A}\right\} $ is
some orthonormal basis of the environment.

Suppose both $\left|c_{0}\right|$ and $\left|c_{1}\right|$ can be
written as square root of rational numbers. Then, we can choose $a$
and $b$ carefully so that

\[
\left|c_{0}\right|=\sqrt{\frac{a}{a+b}},
\]

\[
\left|c_{1}\right|=\sqrt{\frac{b}{a+b}}.
\]

In this case, we reach an equal-weighed superposition of $S$ and
$M$, which is, according to the same reasoning presented above for
a two-state system, envariant with respect with swaps:

\[
\left|\psi\right\rangle _{S+A}=\frac{1}{\sqrt{a+b}}\left\{ e^{i\phi_{0}}\sum_{k=1}^{a}\left|s_{0}\right\rangle _{S}\left|\tilde{m}_{k}\right\rangle _{M}\left|\tilde{a}_{k}\right\rangle _{A}+e^{i\phi_{1}}\sum_{k=a+1}^{a+b}\left|s_{1}\right\rangle _{S}\left|\tilde{m}_{k}\right\rangle _{M}\left|\tilde{a}_{k}\right\rangle _{A}\right\} .
\]

Therefore, we have the same probability of measuring any of the states
$\left|s_{0}\right\rangle _{S}\left|\tilde{m}_{k}\right\rangle _{M}$
or $\left|s_{1}\right\rangle _{S}\left|\tilde{m}_{k}\right\rangle _{M}$.
However, we do not have access to the state of $M$ any more than
we are able to tell the state of the environment $\left|\tilde{a}_{k}\right\rangle _{A}$.
We are only interested on whether we have $\left|s_{0}\right\rangle _{S}$
or $\left|s_{1}\right\rangle _{S}$. Hence, if the probability of
measuring each of the $\left|s_{0}\right\rangle _{S}\left|\tilde{m}_{k}\right\rangle _{M}$
or $\left|s_{1}\right\rangle _{S}\left|\tilde{m}_{k}\right\rangle _{M}$
is the same and equal to:

\[
P\left(\left|s_{0}\right\rangle _{S}\left|\tilde{m}_{k}\right\rangle _{M}\right)=\frac{1}{a+b},
\]
\[
P\left(\left|s_{1}\right\rangle _{S}\left|\tilde{m}_{k}\right\rangle _{M}\right)=\frac{1}{a+b},
\]
we must add all the probabilities for each outcome of $\left|\tilde{m}_{k}\right\rangle _{M}$
to find the total probability of measuring $\left|s_{0}\right\rangle _{S}$
or $\left|s_{1}\right\rangle _{S}$:

\[
P\left(\left|s_{0}\right\rangle _{S}\right)=\sum_{k=1}^{a}P\left(\left|s_{0}\right\rangle _{S}\left|\tilde{m}_{k}\right\rangle _{M}\right)=\frac{a}{a+b}=\left|c_{0}\right|^{2},
\]
\[
P\left(\left|s_{1}\right\rangle _{S}\right)=\sum_{k=a+1}^{a+b}P\left(\left|s_{1}\right\rangle _{S}\left|\tilde{m}_{k}\right\rangle _{M}\right)=\frac{b}{a+b}=\left|c_{1}\right|^{2},
\]
which is Born's law.

In case we cannot represent $\left|c_{0}\right|$ and $\left|c_{1}\right|$
as square roots of rational numbers, we may repeat the argument increasing
$a+b$ as much as we want while keeping

\[
\sqrt{\frac{a}{a+b}}<\left|c_{0}\right|<\sqrt{\frac{a+1}{a+b}},
\]

\[
\sqrt{\frac{b}{a+b}}>\left|c_{1}\right|>\sqrt{\frac{b-1}{a+b}},
\]
thus rendering the gap between the actual coefficients and square
root of the ratios arbitrarily small. Thus we conclude Zurek's argument
derivation of Born's rule via decoherence.

\end{document}